\newif\ifconfver
\confvertrue        

\newif\ifplainver  
\plainvertrue

\newif\ifhide  
\hidetrue

\ifplainver
    \confverfalse   
\fi

\ifconfver
     \documentclass[10pt,twocolumn,twoside]{IEEEtran}
\else
    \ifplainver
        \documentclass[11pt]{article}
        \usepackage{fullpage}
    \else
        \documentclass[12pt,draftcls,onecolumn]{IEEEtran}
    \fi
\fi

\usepackage{etoolbox}%
\usepackage{xpatch}
\usepackage{blindtext}
\usepackage{tocloft}%
\newlength{\articlesectionshift}%
\let\LaTeXStandardSection\section
\let\LaTeXStandardTheSection\thesection
\let\LaTeXStandardTheSubSection\thesubsection
\let\LaTeXStandardTheSubSubSection\thesubsubsection
\let\LaTeXStandardTheParagraph\theparagraph

\makeatletter
\newcounter{titlecounter}

\xpretocmd{\maketitle}{\ifnumgreater{\value{titlecounter}}{1}}{\clearpage}{}{} 
\xpatchcmd{\maketitle}{\let\maketitle\relax\let\@maketitle\relax}{\refstepcounter{titlecounter}\begingroup
  \addtocontents{toc}{\begingroup\addtolength{\cftsecindent}{-\articlesectionshift}}%
  \addcontentsline{toc}{section}{\protect{\numberline{\thetitlecounter}{\@title~ \@author}}}%
  \addtocontents{toc}{\endgroup}
}{%
  \typeout{Patching was successful}
}{%
  \typeout{patching failed}
}%

\def\@IEEEdestroythesectionargument#1{\LaTeXStandardSection{#1}}%

\xapptocmd{\maketitle}{%
\renewcommand{\thesection}{\LaTeXStandardTheSection}%
\renewcommand{\thesubsection}{\LaTeXStandardTheSubSection}%
\renewcommand{\thesubsubsection}{\LaTeXStandardTheSubSubSection}%
\renewcommand{\theparagraph}{\LaTeXStandardTheParagraph}%
}{}{}%

\@addtoreset{section}{titlecounter}

\usepackage{calc,amsfonts,amssymb,amsmath,bm,url,color,theorem,graphicx,cite}
\usepackage{psfrag,float}
\usepackage{algorithm}
\usepackage{algorithmic}
\usepackage{soul}
\usepackage{enumitem}
\usepackage{bbm}
\usepackage{shortcuts_OPT}
\usepackage{multirow}

\usepackage{eqparbox}

\definecolor{orange}{RGB}{255,107,0}

\def\red{\color{red}}

\newtheorem{Fact}{Fact}
\newtheorem{Lemma}{Lemma}

\newtheorem{Theorem}{Theorem}

\newtheorem{Asm}{Assumption}
\theorembodyfont{\rmfamily}

\newtheorem{Remark}{Remark}

\newcommand\bw{\ensuremath{{\bm w}}}

\newcommand\bx{\ensuremath{{\bm x}}}
\newcommand\by{\ensuremath{{\bm y}}}

\newcommand\bz{\ensuremath{{\bm z}}}
\newcommand\bp{\ensuremath{{\bm p}}}

\newcommand\ba{\ensuremath{{\bm a}}}

\newcommand\bA{\ensuremath{{\bm A}}}

\newcommand\bv{\ensuremath{{\bm v}}}

\newcommand\bs{\ensuremath{{\bm s}}}

\usepackage{tikz}
\usetikzlibrary{arrows.meta}
\usetikzlibrary{decorations}
\usetikzlibrary{calc}
\usetikzlibrary{shapes.geometric}
\usetikzlibrary{external}

\begin{document}

\bibliographystyle{IEEEtran}

\newcommand{\papertitle}{
Federated Block Coordinate Descent Scheme for Learning Global and Personalized Models
}

\newcommand{\paperabstract}{
    In federated learning, models are learned from users' data that are held private in their edge devices, by aggregating them in the service provider's ``cloud'' to obtain a global model.
    Such global model is of great commercial value in, e.g., improving the customers' experience. In this paper we focus on two possible areas of improvement of the state of the art.
    First, we take the difference between user habits into account and propose a quadratic penalty-based formulation, for efficient learning of the global model that allows to personalize local models.
    Second, we address the latency issue associated with the heterogeneous training time on edge devices, by exploiting a hierarchical structure modeling communication not only between the cloud and edge devices, but also within the cloud.
    Specifically, we devise a tailored block coordinate descent-based computation scheme, accompanied with communication protocols~for both the synchronous and asynchronous cloud settings.
    We characterize the theoretical convergence rate of the algorithm, and provide a variant that performs empirically better.
    We also prove that the asynchronous protocol, inspired by multi-agent consensus technique,  has the potential for large gains in latency compared to a synchronous setting when the edge-device updates are intermittent.
    Finally, experimental results are provided that corroborate not only the theory, but also show that the system leads to faster convergence for personalized models on the edge devices, compared to the state of the art.
}


\ifplainver


    \title{\papertitle}

    \author{
    Ruiyuan Wu$^\dag$,
    Anna Scaglione$^\ddag$,
    Hoi-To Wai$^\star$,
    Nurullah Karakoc$^\ddag$,\\
    Kari Hreinsson$^\ddag$,
    and Wing-Kin Ma$^\dag$
    \\ ~ \\
    $^\dag$Department of Electronic Engineering, The Chinese University of Hong Kong, \\
    Hong Kong SAR of China \\ ~ \\
    $^\ddag$School of Electrical Computer and Energy Engineering, Arizona State University, USA \\ ~ \\
    $^\star$Department of Systems Engineering and Engineering Management, \\
    The Chinese University of Hong Kong, Hong Kong SAR of China\\
    }

    \maketitle

    \begin{abstract}
    \paperabstract
    \end{abstract}

\else
    \title{\papertitle}

    \ifconfver \else {\linespread{1.1} \rm \fi

    \author{Ruiyuan Wu, Hoi-To Wai, and Wing-Kin Ma
    }

    \maketitle

    \ifconfver \else
        \begin{center} \vspace*{-2\baselineskip}
        \end{center}
    \fi

    \begin{abstract}
    \paperabstract
    \end{abstract}


    \begin{IEEEkeywords}\vspace{-0.0cm}
        Hyperspectral super-resolution, coupled matrix factorization, non-convex optimization
    \end{IEEEkeywords}

    \ifconfver \else \IEEEpeerreviewmaketitle} \fi

 \fi

\ifconfver \else
    \ifplainver \else
        \newpage
\fi \fi

\section{Introduction}
\label{Intro}
Over the past few years, {\it federated learning}, an emerging branch of distributed learning, has attracted increasing attention \cite{mcmahan2016communication,li2018federated,bonawitz2019towards,yang2019federated}.
It focuses on scenarios where users' data are processed for training machine learning models locally, i.e. on users' edge devices such as cell phones and wearable devices, so that the data remain private.
Data privacy is, in fact, a top priority; users are willing to share trained models with reliable service providers, but not necessarily the raw data.
In this context, the service provider seeks a global model---which can, in turn, enhance the performance for the users---by aggregating these local models.
Such global model reflects the ``wisdom of the crowd" and helps the service provider to better understand customer preferences.
What distinguishes federated learning from conventional distributed learning (where the optimization often happens in stable data centers) are the following aspects \cite{li2019federated}:
\begin{enumerate}[leftmargin=0cm,itemindent=.5cm,labelwidth=\itemindent,labelsep=0cm,align=left,label=$\bullet$]
\item
    the heterogeneity of the data and of the computational power available in the users' different edge devices;
\item
    the intermittent (and costly) nature of the communication between edge devices and the cloud.
\end{enumerate}

To explain it in mathematical terms, the following problem is prototypical in federated learning:
\begin{equation}
\label{eq:org_prob}
\begin{aligned}
  \min_{\bx_i,\bz\in\mathcal{X}}~&\sum_{i\in\mathcal{Q}}g_i(\bx_i),~~
  {\rm s.t.}~\bx_i=\bz,~\forall i\in\mathcal{Q},
\end{aligned}
\end{equation}
where $\mathcal{Q}$ is the set of edge devices, $\bx_i$ and $g_i$ are the model parameter and cost function on the $i$th edge device that depend on the local data, and $\mathcal{X}$ is the feasible region.
Specifically, we can write $g_i(\bx_i):=\frac{1}{|\mathcal{S}_i|}\sum_{r\in\mathcal{S}_i}h(\bx_i;\bs_r)$, where $h$ is the training loss function, $\mathcal{S}_i$ the index set of training data on the $i$th edge device, $|\mathcal{S}_i|$ the number of elements in the set $\mathcal{S}_i$, and $\bs_r$ one of such samples.
To solve problem~\eqref{eq:org_prob}, federated learning methods typically adopt a recursive mechanism: edge devices process their own training data to update local models $\bx_i$'s,
and a cloud is introduced to aggregate $\bx_i$'s for a global~model~$\bz$ and synchronize all the local models with the updated $\bz$.
This process is considered standard in the current development \cite{mcmahan2016communication,li2018federated}.
However, we notice that two facts leave some space for improvement:
\begin{enumerate}[leftmargin=0cm,itemindent=.5cm,labelwidth=\itemindent,labelsep=0cm,align=left,label=(\roman*)]
    \item
    {\it Is it efficient to maintain the same model everywhere?}
    Data are generally non-independent~and identically distributed (i.i.d.) on edge devices, since they reflect the usage habits of different users.
    In light of this, edge-device models that work well on data of their users' interest (which are also non-i.i.d.) may suffice---in other words, personalized models may be better at the tasks they are primarily used for.
    When a sole model is maintained, users may sacrifice their customer experience in order to improve the global model that is more beneficial to the service provider to boost new users' models.
    \item
    {\it Is the synchronous cloud model effective?}
    The cloud consists of a cluster of servers that work in parallel, and an edge device only needs to talk with one such cloud server.
    When regarding the cloud as a sole computation resource, one implicitly enforces some form of cloud synchronization, or Sync-cloud (achievable by techniques such as AllReduce~\cite{patarasuk2009bandwidth});
    see Fig.~\ref{fig:architecture}(a) for an illustration.
    In federated learning, this synchronization may lead to significant update latency.
    To see this, recall that due to the capacity heterogeneity of communication and computation, the times of availability and local training time from edge devices can vary significantly.
    Consequently, a possible scenario is that most cloud servers are stranded by a few ``slow'' activated~edge~devices~that~even~never~talk~with~them.
\end{enumerate}
\subsection{Contributions}
The focus of this paper is on addressing the above two issues.
Our contributions and novelty can be summarized as follows:
\begin{enumerate}[leftmargin=0cm,itemindent=.5cm,labelwidth=\itemindent,labelsep=0cm,align=left,label=$\bullet$]
\item
    Our work proposes a tailored hierarchical communication architecture for federated learning.
    This structure, composed of master-slave and multi-agent networks, albeit admitting a ``surprisingly'' familiar look, has not been studied before.
\item
    We propose a lightweight block coordinate descent computation scheme, equipped with judiciously designed communication protocols, which is the first work that can simultaneously achieve the model personalization and cloud server asynchronous updates---two seemly irrelevant issues that are actually closely related to federated learning (c.f. Remark 4).
\item
    The sublinear convergence rate of the proposed algorithm is shown.
    Since our communication architecture contains two layers of information exchange, the analysis requires different analytical tools from the existing work.
\item
    We provide latency analysis to demonstrate the efficacy of the asynchronous update for federated learning.
    {Our latency analytical framework practically allows to estimate runtime from the distribution of the edge-device message arrival time, and to connect the number of cloud servers involved in each update with the runtime, which is new.}
\item
    We carry out reasonable numerical experiments on standard machine learning applications to support our claims.
\end{enumerate}

\subsection{Related Work}
FedAvg~\cite{mcmahan2016communication}, a simple iterative algorithm, is considered the first work in federated learning.
At each round of FedAvg,
the cloud sends the global model to part of the edge devices that are activated;
then, the activated edge devices update its local model by fixed epochs of stochastic gradient descent (SGD) on local cost functions;
finally, the cloud aggregates the uploaded local models (via a weighted summation) as the new global model.
{The majority of follow-up work adopts a similar mechanism as that introduced by FedAvg~\cite{li2019federated}.}
For example, a more recent variant called FedProx~\cite{li2018federated} differs from FedAvg on the edge-device update step, where it imposes an additional proximal term to the cost function, and allows the use of time-varying epochs~of faster algorithms, such as accelerated SGD.

Model personalization is a natural, but sometimes ignored, issue under the federated learning scenarios.
In the work of~\cite{mohri2019agnostic}, the authors argue that, due to personal preferences, models trained via FedAvg may be biased towards the interest of the majority.
They propose AFL that has a sophisticated mechanism to determine the weights of edge-device models involved in the cloud aggregation, instead of the simple data size ratio used in FedAvg.
Per-FedAvg~\cite{fallah2020personalized} distinguishes the global model from the ones on edge devices.
Their goal is to seek a good ``initialization'', as the global model, that can be easily upgraded locally to be optimal for each edge device with a few steps of simple updates.
\cite{smith2017federated} regards model personalization as a multi-task learning problem.
Their proposed MOCHA can learn separate but related models for each device, while leveraging a shared representation via multi-task learning.
{Attesting the importance of model personalization is the work published during the preparation and submission of this paper~\cite{jiang2019improving,arivazhagan2019federated,wu2020personalized,hu2020personalized} that focus exclusively on this issue.}
On the other hand, none of the work considers the penalty-based approach as we do.

{In contrast, there is a vast amount of literature on distributed learning in asynchronous multi-agents' networks; see~\cite{nedic2018network} for a recent survey.
While such implementations require typically more iterations than the master-slave ones, they have no  coordination overhead. Recently~\cite{pmlr-v97-assran19a} demonstrates experimentally that the reduction on overhead yields significant benefits in terms of runtime.
When tasked with a deep-learning problem on a large distributed database the asynchronous multi-agent algorithm runs faster than its master-slave counterpart, because relaxed coordination requirements in turn help complete each update without lags, compared to the  traditional master-slave or incremental architectures.
As of the submission of this paper, we have not seen the study of this topic in the context of federated learning,  as well as no prior theory on how to characterize the performance trends versus the runtime of the algorithm, as opposed to simply focusing on number of iterations required.}
Though there is work mentioning asynchronous federated learning~\cite{chen2019asynchronous,lu2019differentially}, they concentrate on the asynchronous update caused by the communication between edge devices and cloud---the cloud is still regarded as a sole point.

\section{Problem Statement}
\begin{figure}[h]
    \centering
    \includegraphics[width=.75\linewidth]{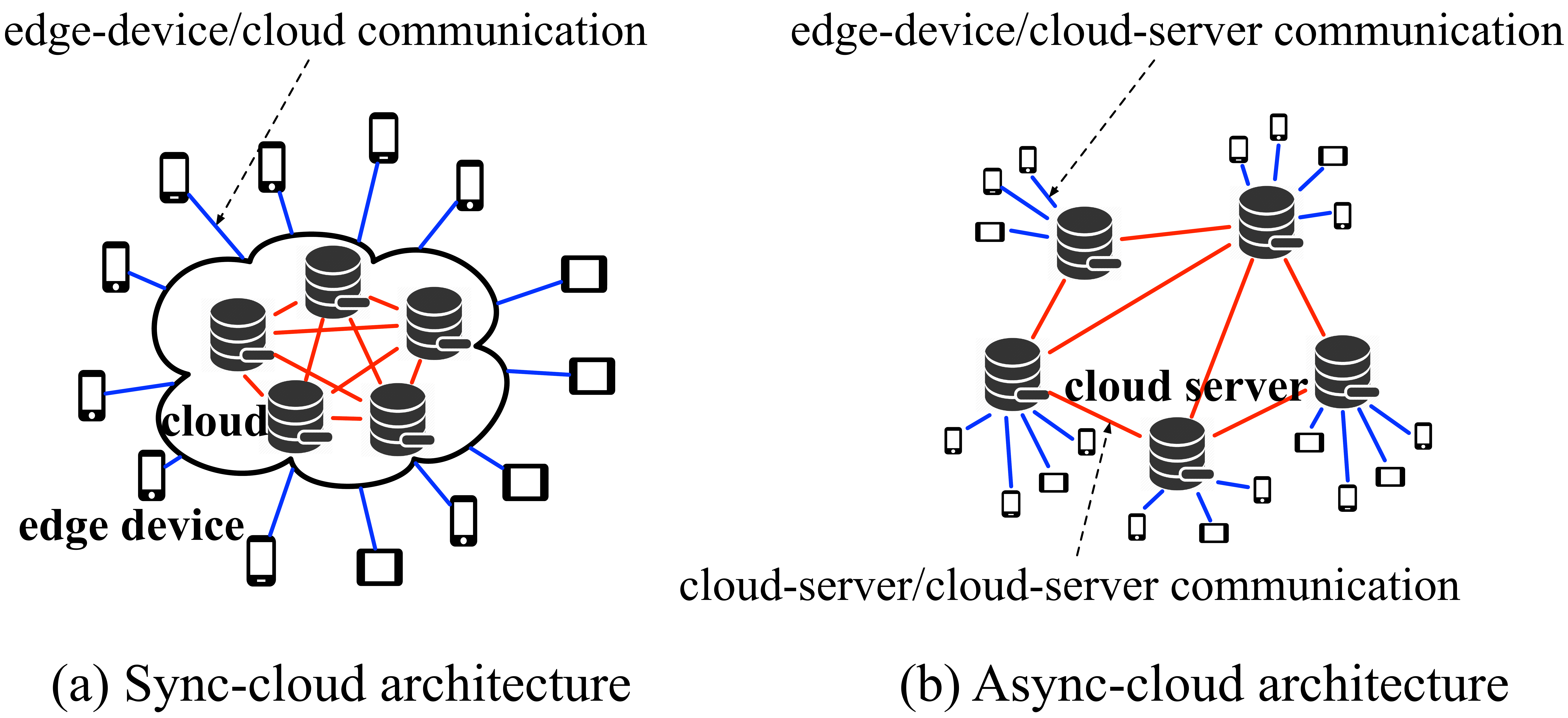}
    \caption{Two communication architectures.}
    \label{fig:architecture}
\end{figure}

We detach the cloud servers from each other and consider both server/server and device/server communication.
To explain, the cloud servers are connected by a (possibly) dynamic multi-agent~graph and they need to talk with each other to achieve consensus, be it exact or asymptotic;
the device/server communication is in a master-slave fashion and, as in general federated learning, intermittent and random.
We call it the Async-cloud architecture, to distinguish it from the Sync-cloud architecture adopted by FedAvg; see Fig.~\ref{fig:architecture} for their difference.
The Async-cloud architecture allows model difference on cloud servers.
As we will show later, such flexibility can be exploited to promote more efficient model updates.

Based on above architectures, our proposed formulation is
\begin{equation}
\label{eq:prob}
\begin{aligned}
    \min_{\substack{\{\bx_i\}_{i\in\mathcal{Q}_n,n\in\mathcal{V}}\\\{\bz_n\}_{n\in\mathcal{V}}}}~&\sum_{n\in\mathcal{V}}\sum_{i\in\mathcal{Q}_n}\left(g_i(\bx_i)+\frac{\gamma_i}{2}\|\bx_i-\bz_n\|^2\right)\\
    {\rm s.t.}\qquad&\bx_i\in\mathcal{X},\forall i\in\mathcal{Q}_n,n\in\mathcal{V},\quad
    \bz_n=\bz_m,~\forall(n,m)\in\mathcal{E},
\end{aligned}
\end{equation}
where $\mathcal{V}$ is the set of cloud servers, $\mathcal{Q}_n$ is the set of edge devices connected to the $n$th cloud server, $\bx_i$'s are edge-device models, $\bz_n$'s are cloud-server models, $\mathcal{E}\subseteq(\mathcal{V}\times\mathcal{V})$ indicates communication between cloud servers, and $\gamma_i>0$ is the penalty parameter.
In the sequel, $\bz_n$'s will be referred to as the global models and $\bx_i$'s the personalized models.
Our formulation distinguishes the global models on different cloud servers, and allows (but penalizes by the quadratic penalty regularizer) the deviations among personalized models.
\begin{Remark}
    The seemingly naive quadratic penalty has recently been revisited in different distributed learning literature.
    In~\cite{zhang2015deep}, this penalty is used to seek better solution in deep learning tasks where the underlying optimization has many local optima.
    This quadratic penalty has also been investigated in adversarial scenarios, where the global model is expected to resist the attack of malicious devices~\cite{yang2020adversary}.
\end{Remark}

\begin{Remark}(Motivation of Global and Personalized Models)
    Consider smartphone keyboard application.
    Users want accurate next-word prediction, which tailored personalized models can better deliver.
    However, each user produces very limited data for training and, thus, its local personalized model may fail to work for new scenarios, which is where the comprehensive knowledge from the global model helps (in our formation, this is obtained by encouraging the personalized model to be close to its global counterpart).
    Also, the global model can serve as an unbiased initialization for new users.
\end{Remark}

\section{Federated Block Coordinate Descent}
In this section, we tackle the problem defined in~\eqref{eq:prob}.
Our development is based on the classic block coordinate descent (BCD) scheme \cite{bertsekas1997nonlinear}:
\begin{subequations}
\label{eq:exBCD}
\begin{align}
\label{eq:client_update}
    &\{\bx_i^{(t+1)}\}_{\substack{i\in\mathcal{Q}_n\\n\in\mathcal{V}}} = \arg\min_{\bx_i\in\mathcal{X}}g_i(\bx_i)+\frac{\gamma_i}{2}\|\bx_i-\bz_n^{(t)}\|^2,\\
\label{eq:server_update}
    &\{\bz_n^{(t+1)}\}_{n\in\mathcal{V}} = \arg\min_{\{\bz_n\}_{n\in\mathcal{V}}}\sum_{n\in\mathcal{V}}\sum_{i\in\mathcal{Q}_n}\frac{\gamma_i}{2}\|\bz_n-\bx_i^{(t+1)}\|^2,\quad {\rm s.t.}\quad\bz_n=\bz_m,~\forall(n,m)\in\mathcal{E}.
\end{align}
\end{subequations}
Our subsequent effort can be interpreted as adapting the BCD update~\eqref{eq:exBCD} to the Sync-cloud and Async-cloud architectures.
Same as FedAvg, we assume that at each round only part of edge devices are available, and use $\mathcal{Q}_n^{(t)}\subseteq\mathcal{Q}_n$ to denote~the set of activated edge devices at round $t$ (i.e., edge devices that are available to do local training and are in stable communication condition).
In the sequel, we will elaborate on the cloud-server and edge-device update separately.

\paragraph*{\bf Update on the edge device.}
If edge device $i$ is not activated, i.e., $i\notin\mathcal{Q}_n^{(t)}$, we have $\bx_i^{(t+1)}=\bx_i^{(t)}$.
Otherwise, time-varying epochs of accelerated stochastic projected gradient (ASPG) is applied.
Specifically, letting $t^-$ be the last round when the $i$th edge device was activated and given the initialization $\bx_i^{(t,0)}=\bx_i^{(t^-,K_i^{(t^-)})}$ and $\bx_i^{(t,-1)}=\bx_i^{(t^-,K_i^{(t^-)}-1)}$, the edge device recursively performs:
\begin{subequations}
\label{eq:AGP}
\begin{align}
    \bx_{i,\sf ex}^{(t,k)} &= \bx_i^{(t,k)}+\zeta (\bx_i^{(t,k)}-\bx_i^{(t,k-1)}),\\
    \bx_i^{(t,k+1)}&=\Pi_{\mathcal{X}}\left(\bx_{i,\sf ex}^{(t,k)}-\eta_x\left(\nabla \tilde{g}_i(\bx_{i,\sf ex}^{(t,k)})+\gamma_i(\bx_{i,\sf ex}^{(t,k)}-\bz_n^{(t)})\right)\right),
\end{align}
\end{subequations}
for $k=0,\ldots,K_i^{(t)}-1$,
where $\eta_x$ is the stepsize, $\zeta \geq 0$ is the momentum~weight,~$\Pi_{\mathcal{X}}$~is the projection onto $\mathcal{X}$, and $\nabla\tilde{g}_i$ is the mini-batch stochastic gradient such that $\nabla\tilde{g}_i = \frac{1}{R}\sum_{r=1}^R\nabla h(\bx_i;\bxi_r^{(t,k)})$
with $\bxi_r^{(t,k)}$ being a sample from the $i$th edge device and $R$ being the batch size.
The final update is $\bx_i^{(t+1)}=\bx_i^{(t,K_i^{(t)})}$.
If $\zeta =0$, update~\eqref{eq:AGP} narrows down to the standard SPG.
Empirically, the ASPG update is observed to converge faster~\cite{beck2009fast}.

\paragraph*{\bf Update on the cloud server.}
We derive the update rule for~\eqref{eq:server_update} for both Sync-cloud and Async-cloud architectures.
In the synchronous case, same as FedAvg, we can denote the global models on all participating cloud servers as $\bz_n=\bz$ for $n\in\mathcal{V}$.
Then, given initialization $\bz^{(t)}$, the cloud updates
\begin{align}
\label{eq:synchronous_update}
    \raisebox{0\normalbaselineskip}{\rm~(Sync-cloud)}\qquad
    \bz^{(t+1)} = \bz^{(t)}-\eta_z\sum_{n\in\mathcal{V}}\sum_{i\in\mathcal{Q}_n}\gamma_i(\bz^{(t)}-\bx_i^{(t+1)}),
\end{align}
where $\eta_z$ is the stepsize.

In the asynchronous setting, we propose the use of the distributed gradient descent (DGD) algorithm~\cite{ram2010distributed}.
Similarly, given initialization $\bz_n^{(t)}$, for $n\in\mathcal{V}$, each cloud server runs
\noindent
\begin{subequations}
\label{eq:DGD}
\begin{align}
    \raisebox{-.0\normalbaselineskip}{\rm~(Async-cloud)}\qquad
    \bw_n^{(t)}=&\sum_{m\in\mathcal{V}}a_{n,m}^{(t)}\bz_m^{(t)},\\
    \bz_n^{(t+1)} =&~\bw_n^{(t)}-\eta_z\sum_{i\in\mathcal{Q}_n}
    \gamma_i( \bw_n^{(t)} -\bx_i^{(t+1)}),
\end{align}
\end{subequations}
where $a_{n,m}^{(t)}$ is a mixing coefficient decided by the server/server communication link at round $t$.

Combining~\eqref{eq:AGP}-\eqref{eq:DGD},
we obtain our FedBCD scheme, which is summarized in Algorithm~\ref{alg:FedBCD_full}.
In Section~\ref{Sec:Anal}, we will characterize~the theoretical convergence guarantee of FedBCD.
\begin{algorithm}[!t]
    \caption{FedBCD for problem~\eqref{eq:prob}}
    \label{alg:FedBCD_full}
    \footnotesize
\begin{algorithmic}[1]
    \STATE {\bfseries input:} for $i\in\mathcal{Q}_n,n,m\in\mathcal{V},t=0,1,\ldots,T-1$:
    model initialization $\bx_i^{(-1,0)}=\bx_i^{(-1,-1)}=\bz_n^{(0)}=\bx^{(0)}\in\mathcal{X}$;
    regularization parameter $\gamma_i$;
    mixing coefficient $a_{n,m}^{(t)}$;
    stepsize $\eta_x$ and $\eta_z$;
    momentum weight $\zeta $;
    cloud server update iteration $K_z$;
    edge device update iteration $K_i^{(t)}\in[K_x]$, with $K_i^{(-1)}=0$ and $K_i^{(t)}=1$ if $i\notin\mathcal{Q}_n^{(t)}$;
    batch size number $R$;
    \\ \vspace*{-.5em}
    \noindent\rule{\linewidth}{.4pt} \vspace*{-1em}
    \STATE {\bf for $t=0,1,\ldots,T-1$ do}
        \STATE \quad{randomly select a subset $\mathcal{Q}_n^{(t)}\subseteq\mathcal{Q}_n,~\forall n\in\mathcal{V}$};\newline
            {\color{white}.}~  {\red \it\% edge devices execute:}
        \STATE \quad{\bf for $i\in\mathcal{Q}_n^{(t)},n\in\mathcal{V}$ do}
            \STATE \qquad $\bx_i^{(t,0)}=\bx_i^{(t^-,K_i^{t^-})}$,
            $\bx_i^{(t,-1)}=\bx_i^{(t^-,K_i^{t^-}-1)}$, where $t^-<t$ is the last round such that $i\in\mathcal{Q}_n^{(t^-)}$;
            \STATE \qquad {\bf for $k=0,\ldots,K_i^{(t)}-1$ do}
                \STATE \quad\qquad randomly draw $R$ samples $\bxi_{i,r}^{(t,k)}$'s stored on the $i$th edge device;\label{alg:apg_3}
                \STATE \quad\qquad $\bx_{i,\sf ex}^{(t,k)} = \bx_i^{(t,k)}+\zeta (\bx_i^{(t,k)}-\bx_i^{(t,k-1)})$;\label{alg:apg_1}
                \STATE \quad\qquad $\bx_{i}^{(t,k+1)}=\Pi_{\mathcal{X}}(\bx_{i,\sf ex}^{(t,k)}-\eta_x(\frac{1}{R}\sum_{r=1}^R\nabla h(\bx_{i,\sf ex}^{(t,k)};\bxi_{i,r}^{(t,k)})+\gamma_i(\bx_{i,\rm ex}^{(t,k)}-\bz_n^{(t)})))$;  \label{alg:apg_2}
            \STATE \qquad {\bf end for}
            \STATE \qquad $\bx_i^{(t+1)}=\bx_i^{(t,K_i^{(t)})}$;
        \STATE \quad{\bf end for}
        \STATE \quad{\bf for $i\notin\mathcal{Q}_n^{(t)},n\in\mathcal{V}$ do}
            \STATE \qquad $\bx_i^{(t+1)}=\bx_i^{(t)}$;
        \STATE \quad{\bf end for}\newline
            {\color{white}...}  {\it \red \% cloud servers execute:}
            \STATE \quad {$\{\bz_n^{(t,0)}\}_{n\in\mathcal{V}}=\{\bz_n^{(t)}\}_{n\in\mathcal{V}}$;}
            \STATE \quad {\bf for $k=0,\ldots,K_z-1$}
            \STATE \qquad {\bf for $n\in\mathcal{V}$}
            \STATE \qquad\quad $\bw_n^{(t,k)}=\sum_{m\in\mathcal{V}}a_{n,m}^{(t)}\bz_m^{(t,k)}$; \label{alg:ca_1}
            \STATE \qquad\quad $\bz_n^{(t,k+1)}=\bw_n^{(t,k)}$$-\eta_z\sum_{i\in\mathcal{Q}_n}\gamma_i(\bw_n^{(t,k)}-\bx_i^{(t+1)})$; \label{alg:ca_2}
            \STATE \qquad {\bf end for}
        \STATE \quad{\bf end for}
        \STATE \quad {$\{\bz_{n}^{(t+1)}\}_{n\in\mathcal{V}}=\{\bz_n^{(t,K_z)}\}_{n\in\mathcal{V}}$;}
   \STATE {\bf end for}\\ \vspace*{-.5em}
   \noindent\rule{\linewidth}{.4pt} \vspace*{-1em}
   \STATE {\bf output: $\{\bx_i^{(T)}\}_{i\in\mathcal{Q}_n,n\in\mathcal{V}},\{\bz_n^{(T)}\}_{n\in\mathcal{V}}$}.
\end{algorithmic}
\end{algorithm}
Note that Algorithm~\ref{alg:FedBCD_full} covers both the Sync-cloud and Async-cloud updates.
For the Sync-cloud update, it is equivalent to fix $|\mathcal{V}|=1$ (though there are multiple cloud servers in practice, they are regarded as a sole point in the algorithm under the Sync-cloud setting).
Also, here we provide a more general version by allowing the cloud servers to run multiple epochs of DGD updates.

\subsection{Sync-Cloud and Async-Cloud Communication Protocols for FedBCD}
\label{sec:protocol}
\begin{figure}[h]
    \centering
    \includegraphics[width=.9\linewidth]{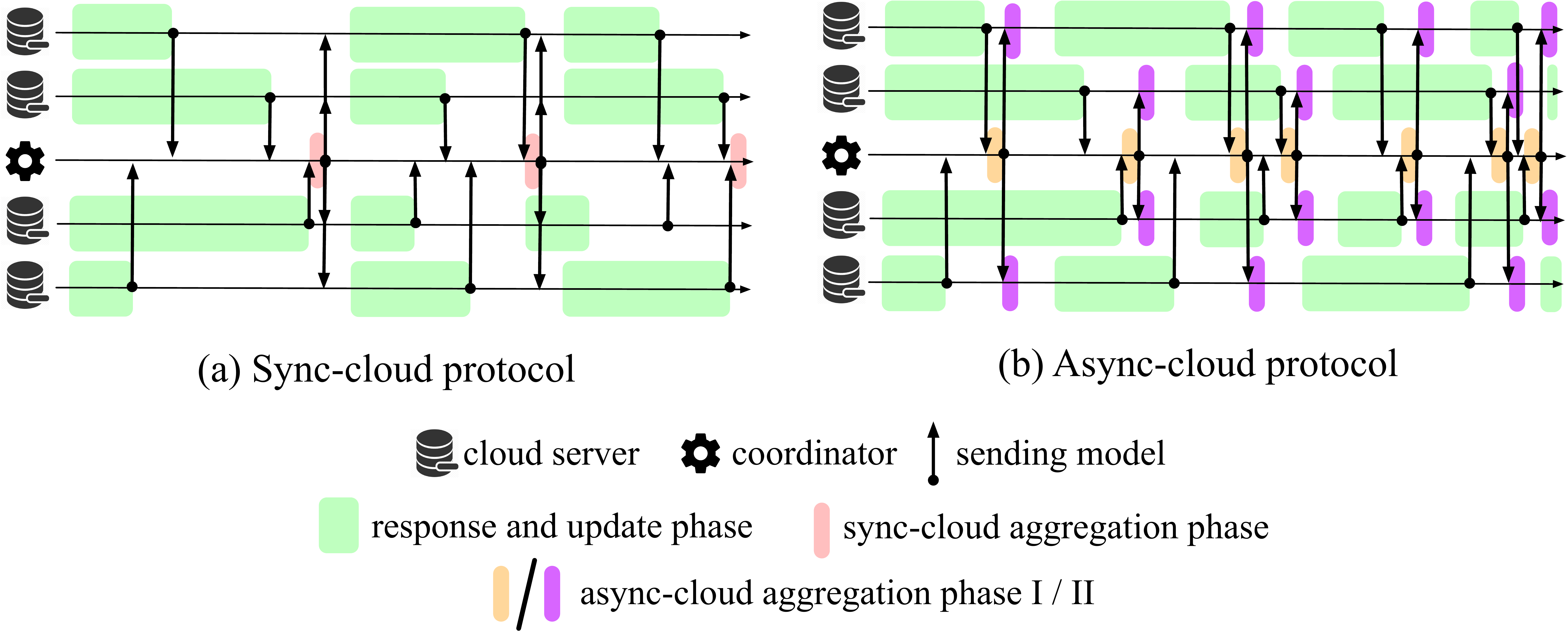}
    \caption{Protocols under Sync-/Async-cloud architectures.}
    \label{fig:protocol}
\end{figure}

We introduce two communication protocols to clarify how FedBCD can be applied, and how the Async-cloud protocol can allow potentially~more~efficient updates.
\begin{enumerate}[leftmargin=0cm,itemindent=.5cm,labelwidth=\itemindent,labelsep=0cm,align=left,label=$\bullet$]
\item
    {\it Sync-cloud protocol.}
    This protocol is similar to that in~\cite{bonawitz2019towards} and has two phases:
    (i) A new round~$t$ starts with the ``response and update'' phase.
    In this phase, cloud servers accept update requests~from edge devices, send the global model $\bz^{(t)}$ to these activated edge devices, and wait for them to~perform the local update~\eqref{eq:AGP}.
    For each cloud server, this phase ends when it receives $\bx_i^{(t+1)}$'s from a pre-determined number of edge devices.
    Then, this cloud server replaces its local copy of $\bx_i^{(t)}$'s with $\bx_i^{(t+1)}$'s (for inactivated edge devices, it simply puts $\bx_i^{(t+1)}=\bx_i^{(t)}$) and sends $\sum_{i\in\mathcal{Q}_n}\bx_i^{(t+1)}$ to a coordinator that keeps the latest global model $\bz^{(t)}$.
    (ii) Once hearing from all cloud servers, this coordinator enters the ``aggregation'' phase,
    where it executes~\eqref{eq:synchronous_update} to obtain $\bz^{(t+1)}$~and broadcasts it back.
    After receiving $\bz^{(t+1)}$, the cloud servers become available again for the next round of update.
\item
    {\it Async-cloud protocol.}
    This protocol requires more careful implementation.
    (i) At round $t$,~cloud servers that finished the ``response and update'' phase send their model $\bz_n^{(t)}$'s to the coordinator.
    (ii) The coordinator waits until it hears from the first $B$ cloud~servers~$\mathcal{V}^{(t)}$~[in Fig.~\ref{fig:protocol}(b) it is $|\mathcal{V}^{(t)}|\!=\!B\!=\!2$], and enters ``aggregation'' phase I.
    Then, this coordinator calculates $\bw^{(t)}=\sum_{n\in\mathcal{V}^{(t)}}\bz_n^{(t)}/B$ and~sends $\bw^{(t)}$ to cloud servers in $\mathcal{V}^{(t)}$.
    (iii) The cloud servers in $\mathcal{V}^{(t)}$ enter ``aggregation'' phase~II, where they use received $\bw^{(t)}$ as $\bw_n^{(t)}$ and update $\bz_n^{(t+1)}$ by~\eqref{eq:DGD};
    the other cloud servers simply~have~$\bz_n^{(t+1)}=\bz_n^{(t)}$.
    Back to our FedBCD scheme, this protocol corresponds to the following setting: $\ba_{n,m}^{(t)}=1/B$ for $n,m\in\mathcal{V}^{(t)}$,
    $\ba_{n,n}^{(t)}=1$ and $\ba_{n,m}^{(t)}=0$ for $n,m\notin\mathcal{V}^{(t)}$;
    $\mathcal{Q}_n^{(t)}=\emptyset$ and $\eta_{z_n}^{(t)}=0$ for $n\notin\mathcal{V}^{(t)}$.
\end{enumerate}

In the Sync-cloud protocol, the model aggregation happens when all cloud servers finish the ``response and update'' phase;
while in the second protocol, cloud servers work asynchronously, and the model aggregation is carried out in a ``first come first serve'' manner.
By constructing the neighborhoods of cloud servers that mix their model dynamically based on those who are ready first, the Async-cloud protocol, as analyzed below, is expected to provide more efficient update;
see Fig.~\ref{fig:protocol} for an illustration.
\begin{Remark}
    Some readers may argue that the use of the coordinator in both protocols is not favorable for multi-agent communication.
    However, unlike in the conventional setting, herein the coordinator only involves lightweight computations.
    We believe that with the advent of software-defined networking, it is fairly natural to assume that specialized servers communications, such as multi-agent message exchanges, can be orchestrated in an effective way to support specific applications.
    Methods that completely eliminate coordination require re-normalizing the weights~\cite{pmlr-v97-assran19a}; typically the algorithms take longer to converge.
\end{Remark}

\paragraph{Latency analysis.}
Here we provide a methodology to gain analytical insight, by assuming a certain distribution for the inter-update time.
Let {\small $\tau_{n}^{(t)}, n\in {\cal V}$} be random time taken by cloud server $n$ to accrue {\small ${\cal Q}_n^{(t)}$} edge-devices updates; assume {\small $\tau_{n}^{(t)}$} are i.i.d. with mean and standard deviation $\mu_{\tau},\sigma_{\tau}$ respectively. Each  {\small $\tau_{n}^{(t)}$} equals the sum of update request arrival time and the computation time associated to {\small ${\cal Q}_n^{(t)}$} edge-devices.
With {\small $|{\cal V}|=N$},  denote by {\small $\tau_{(k)}^{(t)},~k=1,\ldots,N$} the ordered statistics of the samples {\small $\tau_{n}^{(t)}$} in increasing order, i.e. {\footnotesize $\tau_{(1)}^{(t)}\leq \tau_{(2)}^{(t)}\leq \ldots\leq \tau_{(N)}^{(t)}=\max_{n\in \mathcal{V}} \tau_{n}^{(t)}$}. Let the latency of the {Sync-cloud} and {Async-cloud} protocols at round $t$ be {\small $\tau_{\mathrm{SC}}^{(t)}$} and  {\small $\tau_{\mathrm{AC}}^{(t)}$}. It follows that {\small $\tau_{\mathrm{SC}}^{(t)}= \tau_{(N)}^{(t)}$} and {\small $\tau_{\mathrm{AC}}^{(t)}=\tau_{(B)}^{(t)}$}. Hence the reduction in the average duration per round is:
\begin{equation}
\label{eq:latency_anal_1}
r_{B,N}:= \mathbb{E} [\tau_{(B)}^{(t)}]/\mathbb{E}[\tau_{(N)}^{(t)} ]
\equiv \mathbb{E}[\tau_{\mathrm{AC}}^{(t)}]/\mathbb{E}[\tau_{\mathrm{SC}}^{(t)}].
\end{equation}
Let $\beta=B/N$ denote the percentage of network nodes involved in each update, and consider the fact that $r_{B,N}\leq r_{B,N-1}$. For large $N$ we can leverage a classic result~\cite{mosteller2006some}, showing the asymptotic normality of ordered statistics for large sample size. In particular by denoting $F_{\tau_n}^{-1}(u)$ the quantile function of the random variable $\tau_n$, we have
\begin{align}
\label{eq:latency_anal_2}
&\tau_{(\beta N)}^{(t)}\sim {\cal N}\left(F_{\tau_n}^{-1}(\beta),
\frac{\beta(1-\beta)}{N\left[f_{\tau_n}\left(F_{\tau_n}^{-1}(\beta)\right)\right]^2}
\right)~~\Rightarrow\quad r_{\beta N,N}
 \lessapprox \frac{F_{\tau_n}^{-1}(\beta)}{F_{\tau_n}^{-1}(1-1/N)},~{\rm for}~N\gg 1.
\end{align}
For $\tau_n^{(t)}$ with sample space $[0,+\infty)$,  {\small $F_{\tau_n}^{-1}(1-1/N)\rightarrow +\infty$}, there is an infinite gain in latency asymptotically. For typical distributions, though, the rate is slow.
For example, for {\small $\tau_n^{(t)}$} following a Weibull distribution, with shape parameter $k$ (irrespective of the scale parameter $\lambda$), we have $r_{\beta N,N}\lessapprox -(\ln(1-\beta))^{\frac 1 k}(\log(N))^{-\frac 1 k}$.

For a random time distribution with finite support $[0,\overline{\tau}]$ with $\overline{\tau}<+\infty$, instead, the reduction saturates to $F_{\tau_n}^{-1}(\beta)/\overline{\tau}$.
Also notice that, in general, the smaller is the percentage of network nodes in each update $\beta$ the slower is the update progress per round, yielding a trade-off between reducing latency and increasing the update progress, which results in an optimum choice of $\beta$.
Finally, it should be pointed out that we only provide a straightforward solution to illustrate the potential of the Async-cloud architecture.
Multi-agent consensus techniques such as DGD have been well studied in distributed learning~\cite{pmlr-v97-assran19a}, opening the door to further advances to promote more~efficient~updates.

\begin{Remark}
    Model personalization and asynchronous updates may seem to be two independent topics at first glance.
    However, we emphasize that both of them inherently match the heterogeneous nature of federated learning---they go hand in hand, due to the idiosyncratic behavior of the edge devices manifesting itself in both the non i.i.d. data and the non-uniform times at which updates are carried out. In view of this, only forcing equal models but applying asynchronous updates (or the reverse) is actually a half-measure.
\end{Remark}

\subsection{An Intuitive Variant of FedBCD}
\label{sec:variant}
In FedBCD, edge devices update local models only when they are in stable communication condition, which could lead to slow local model update.
On the other hand, it is reasonable to assume that edge devices can also train local models offline.
For instance, think about a phone that is charged and idle.
This is a good time for the phone to do local training, even if it is not connected to Wi-Fi.
Thus, we have $\mathcal{Q}_n^{(t)}\subseteq\tilde{\mathcal{Q}}_n^{(t)}\subseteq\mathcal{Q}_n$, where $\tilde{\mathcal{Q}}_n^{(t)}$ is the set of edge devices that are available for local training but in bad communication condition.
We can then take the intuitive approach suggested in~\cite{zhang2015deep} by tackling the cost function and quadratic penalty in~\eqref{eq:prob} separately.
Simply speaking, at round $t$, edge devices in~$\tilde{\mathcal{Q}}_n^{(t)}$ run local~training using cost functions $g_i$'s.
Then, the subset of edge devices in $\mathcal{Q}_n^{(t)}$ sends update requests to the cloud for the global model to adjust their local models.
The cloud-server updates remain unchanged.
We call the resulting scheme FedBCD-I, and summarize it in Algorithm~\ref{alg:FedBCD_I_full}.
In a nutshell, the spirit is to allow edge devices to carry out independent update locally based on the local training cost function, and use the quadratic penalty with the global model received intermittently to adjust their local models.
\begin{algorithm}[!t]
    \caption{FedBCD-I for problem~\eqref{eq:prob}}
    \label{alg:FedBCD_I_full}
    \footnotesize
\begin{algorithmic}[1]
    \STATE {\bfseries input:} for $i\in\mathcal{Q}_n,n,m\in\mathcal{V},t=0,1,\ldots,T-1$:
    model initialization $\bx_i^{(-1,0)}=\bx_i^{(-1,-1)}=\bz_n^{(0)}=\bx^{(0)}\in\mathcal{X}$;
    regularization parameter $\gamma_i$;
    mixing coefficient $a_{n,m}^{(t)}$;
    stepsize $\eta_x$ and $\eta_z$;
    momentum weight $\zeta $;
    cloud server update iteration $K_z$;
    edge device update iteration $K_i^{(t)}\in[K_x]$, with $K_i^{(-1)}=0$ and $K_i^{(t)}=1$ if $i\notin\mathcal{Q}_n^{(t)}$;
    batch size number $R$;
    \\ \vspace*{-.5em}
    \noindent\rule{\linewidth}{.4pt} \vspace*{-1em}
    \STATE {\bf for $t=0,1,\ldots,T-1$ do}
        \STATE \quad{randomly select a subset $\mathcal{Q}_n^{(t)}\subseteq\tilde{\mathcal{Q}}_n^{(t)}\subseteq\mathcal{Q}_n,~\forall n\in\mathcal{V}$};\newline
            {\color{white}.}~  {\red \it\% edge devices execute:}
        \STATE \quad{\bf for $i\in\tilde{\mathcal{Q}}_n,n\in\mathcal{V}$ do}
            \STATE \qquad $\bx_i^{(t,0)}=\bx_i^{(t^-,K_i^{t^-})}$,
            $\bx_i^{(t,-1)}=\bx_i^{(t^-,K_i^{t^-}-1)}$, where $t^-<t$ is the last round such that $i\in\mathcal{Q}_n^{(t^-)}$;
            \STATE \qquad {\bf for $k=0,\ldots,K_i^{(t)}-1$ do}
                \STATE \quad\qquad randomly draw $R$ samples $\bxi_{i,r}^{(t,k)}$'s stored on the $i$th edge device;\label{alg:apg_3}
                \STATE \quad\qquad $\bx_{i,\sf ex}^{(t,k)} = \bx_i^{(t,k)}+\zeta (\bx_i^{(t,k)}-\bx_i^{(t,k-1)})$;\label{alg:apg_1}
                \STATE \quad\qquad $\bx_{i}^{(t,k+1)}=\Pi_{\mathcal{X}}(\bx_{i,\sf ex}^{(t,k)}-\frac{\eta_x}{R}\sum_{r=1}^R\nabla h(\bx_{i,\sf ex}^{(t,k)};\bxi_{i,r}^{(t,k)}))$;  \label{alg:apg_2}
            \STATE \qquad {\bf end for}
            \STATE \qquad $\bx_i^{(t+1)}=\bx_i^{(t,K_i^{(t)})}$;
        \STATE \quad{\bf end for}
        \STATE \quad{\bf for $i\in\mathcal{Q}_n^{(t)},n\in\mathcal{V}$ do}
            \STATE \qquad $\bx_i^{(t,0)} = \bx_i^{(t+1)}$;
            \STATE \qquad {\bf for $k=0,\ldots,K_i^{(t)}-1$ do}
            \STATE \qquad\quad $\bx_i^{(t,k+1)}=\Pi_{\mathcal{X}}(\bx_i^{(t,k)}-\gamma_i(\bx_i^{(t,k)}-\bz_n^{(t)}))$;
            \STATE \qquad {\bf end for}
            \STATE \qquad $\bx_i^{(t+1)}=\bx_i^{(t,K_i^{(t)})}$
        \STATE \quad{\bf end for}\newline
            {\color{white}...}  {\it \red \% cloud servers execute:}
            \STATE \quad {$\{\bz_n^{(t,0)}\}_{n\in\mathcal{V}}=\{\bz_n^{(t)}\}_{n\in\mathcal{V}}$;}
            \STATE \quad {\bf for $k=0,\ldots,K_z-1$}
            \STATE \qquad {\bf for $n\in\mathcal{V}$}
            \STATE \qquad\quad $\bw_n^{(t,k)}=\sum_{m\in\mathcal{V}}a_{n,m}^{(t)}\bz_m^{(t,k)}$; \label{alg:ca_1}
            \STATE \qquad\quad $\bz_n^{(t,k+1)}=\bw_n^{(t,k)}$$-\eta_z\sum_{i\in\mathcal{Q}_n^{(t)}}\gamma_i(\bw_n^{(t,k)}-\bx_i^{(t+1)})$; \label{alg:ca_2}
            \STATE \qquad {\bf end for}
        \STATE \quad{\bf end for}
        \STATE \quad {$\{\bz_{n}^{(t+1)}\}_{n\in\mathcal{V}}=\{\bz_n^{(t,K_z)}\}_{n\in\mathcal{V}}$;}
   \STATE {\bf end for}\\ \vspace*{-.5em}
   \noindent\rule{\linewidth}{.4pt} \vspace*{-1em}
   \STATE {\bf output: $\{\bx_i^{(T)}\}_{i\in\mathcal{Q}_n,n\in\mathcal{V}},\{\bz_n^{(T)}\}_{n\in\mathcal{V}}$}.
\end{algorithmic}
\end{algorithm}



We can implement FedBCD-I under the same communication protocols introduced for FedBCD.
The difference is that now edge devices are allowed to run local training offline, and thus we need to adapt the ``response and update'' phase to this setting.
Take the Sync-cloud protocol for example:
\paragraph*{\it $\bullet$ Sync-cloud protocol (for FedBCD-I).}
(i) At a round $t$, edge devices that are available run local training (for a round) offline based on lines~4-12 in Algorithm~\ref{alg:FedBCD_I_full}.
(ii) Among them, edge devices that are also in temporarily stable communication condition send their update requests to the cloud servers in order to join the ``response and update'' phase.
Edge devices that miss this phase will start the next round of local training and seek to join it at the next round.
If an edge device has run a fixed rounds of updates without communication with the cloud, it will stop training until it successfully gets activated.
(iii)
During the ``response and update'' phase, cloud servers accept update requests~from edge devices, send the global model $\bz^{(t)}$ to these activated edge devices, and wait for them to~adjust their local models based on lines~13-19 in Algorithm~\ref{alg:FedBCD_I_full}.
For each cloud server, this phase ends when it receives $\bx_i^{(t+1)}$'s from a pre-determined number of edge devices.
Then, this cloud server replaces its local copy of $\bx_i^{(t)}$'s with $\bx_i^{(t+1)}$'s and sends $\sum_{i\in\mathcal{Q}_n^{(t)}}\bx_i^{(t+1)}$ to a coordinator that keeps the latest global model $\bz^{(t)}$, where $\mathcal{Q}_n^{(t)}$ is the set of activated edge devices.
(iii)
Once hearing from all cloud servers, this coordinator enters the ``aggregation'' phase,
where it executes lines 21-26 to obtain $\bz^{(t+1)}$~and broadcasts it back.
After receiving $\bz^{(t+1)}$, the cloud servers become available again for the next round of update.

The Async-cloud protocol is almost the same as that for FedBCD, with the same change made on the details of the ``check-in'' and ``edge-device update'' phases.
We omit the details to avoid redundancy.

\section{Theoretical Convergence Analysis}
\label{Sec:Anal}
We investigate the behavior of FedBCD in this section.
Our result applies to both the Sync-cloud and Async-cloud settings.
Let us start with the following assumptions:
\begin{Asm} \label{Asm:subset}
    The server/server and device/server communication satisfies that
    \begin{enumerate}[leftmargin=0cm,itemindent=.5cm,labelwidth=\itemindent, nolistsep,labelsep=0cm,align=left,label=(\roman*)]
    \item
        Each edge device is activated at least once every $p$ iterations, where $p<\infty$.
    \item
        In the Async-cloud update~\eqref{eq:DGD}, the communication graph within cloud servers is possibly time-varying and satisfies, for some $q<\infty$ that, {\small $(\mathcal{V},\bigcup_{t=1,\ldots,q}\mathcal{E}^{(t_0+t)})$} is strongly connected for~all~$t_0$.
    \end{enumerate}
\end{Asm}
\begin{Asm}
\label{Asm:function} In
    Problem~\eqref{eq:prob} the conditions below are met:
    \begin{enumerate}[leftmargin=0cm,itemindent=.5cm,labelwidth=\itemindent, nolistsep,labelsep=0cm,align=left,label=(\roman*)]
        \item
        \label{Asm:1}
        The feasible set $\mathcal{X}$ is convex and compact.
        \item
        \label{Asm:3}
        For all $i\in\mathcal{Q}_n,n\in\mathcal{V}$, function $g_i$ is bounded on $\mathcal{X}$,
        and is $L_i$-Lipschitz smooth on $\tilde{\mathcal{X}}$, where {\small $\tilde{\mathcal{X}} = {\rm conv}[\bigcup_{0\leq\zeta\leq1}\{\bx+\zeta(\bx-\by)\mid|\bx,\by\in\mathcal{X}\}]$} is the feasible set extended by the momentum update.
        Also,~its gradient $\nabla g_i$ is bounded on $\mathcal{X}$.
    \end{enumerate}
\end{Asm}
\begin{Asm} \label{Asm:5}
The mixing coefficient {\small $\{a_{n,m}^{(t)}\}_{t=0,1,\ldots,}$} in Async-cloud update~\eqref{eq:DGD} satisfies the conditions below:
\begin{enumerate}[leftmargin=0cm,itemindent=.5cm,labelwidth=\itemindent, nolistsep,labelsep=0cm,align=left,label=(\roman*)]
    \item
    There exists a scalar {\small $c>0$} such that {\small $a_{n,m}^{(t)}\geq c$} if {\small $(n,m)\in\mathcal{E}^{(t)}$} and {\small $a_{n,m}^{(t)}=0$} otherwise.
    \item
    (doubly stochastic) {\small $\sum_{m\in\mathcal{V}}a_{n,m}^{(t)}=\sum_{n\in\mathcal{V}}a_{n,m}^{(t)}=1$}.
\end{enumerate}
\end{Asm}
\begin{Asm}
\label{asm:stepsize}
The (constant) stepsizes and momentum weight satisfy
\begin{enumerate}[leftmargin=0cm,itemindent=.5cm,labelwidth=\itemindent, nolistsep,labelsep=0cm,align=left,label=(\roman*)]
    \item
    $\eta_{x}\leq \min_{i\in\mathcal{Q}_n,n\in\mathcal{V}}\{1/(L_i+\gamma_i)\}$.
    \item
    $\eta_z=d/\sqrt{T}\leq1/(\max_{i\in\mathcal{Q}_n,n\in\mathcal{V}}\{\gamma_i\}\cdot\max_{n\in\mathcal{V}}\{|\mathcal{Q}_n|\})$,
    for a positive constant $d$ and total communication round $T$.
    \item
    $\zeta  \leq \min_{i\in\mathcal{Q}_n,n\in\mathcal{V}}\{\omega/\sqrt{1+\rho_i\eta_x}\}$,
    for some constants $\omega\in(0,1)$ and $\rho_i$ such that $\rho_i\leq L_i+\gamma_i$.
\end{enumerate}
\end{Asm}

\begin{Asm}
\label{asm:batch_size}
    For any $i$, let $\nabla h(\bx_i;\bxi_{i})$ be a stochastic gradient of $g_i(\bx_i)$, where $\bxi_{i}$ is a random sample from the $i$th edge device
    [recall that in~\eqref{eq:org_prob} we denote {\small $g_i(\bx_i):=\frac{1}{|\mathcal{S}_i|}\sum_{r\in\mathcal{S}_i}h(\bx_i;\bs_r)$}].
    It holds for any $\bx_i$ and $\bxi_i$ that {\small $\mathbb{E}[\nabla g_i(\bx_i)-\nabla h(\bx_i;\bxi_i)]=\bm0$} and  {\small $\mathbb{E}[\|\nabla g_i(\bx_i)-\nabla h(\bx_i;\bxi_i)\|^2]\leq\sigma^2$}.
\end{Asm}
Let us briefly examine our assumptions:
Assumption~\ref{Asm:subset} makes sure that no edge device or cloud server is~isolated;
Assumptions~\ref{Asm:function}-\ref{Asm:5} are common in constrained optimization and average consensus algorithms;
Assumption~\ref{asm:stepsize} requires that the stepsize and the momentum weight are bounded;
Assumption~\ref{asm:batch_size} is typical for stochastic gradient descent.
Careful readers may notice that the Async-cloud protocol in Section~\ref{sec:protocol} violates~Assumption~\ref{asm:stepsize} by considering time-varying $\eta_z$.
This issue can be fixed by minor modification; the resulting version is, however, more complicated in terms of implementation and is detailed in Appendix~\ref{appendix:1}.

We characterize the convergence of FedBCD to a stationary point.
Our convergence metric is the constrained version of the one in~\cite{pmlr-v97-assran19a}.
\begin{Theorem}
\label{thm}
Suppose that Assumptions~\ref{Asm:subset}-\ref{asm:batch_size} hold, communication round $T$ is sufficiently large (see the supplementary document for a clear definition), and batch size $R$ is used to calculate the stochastic gradients (see the detailed discussion of the stochastic gradient below~\eqref{eq:AGP})
\begin{enumerate}[leftmargin=0cm,itemindent=.5cm,labelwidth=\itemindent, nolistsep,labelsep=0cm,align=left,label=(\roman*)]
    \item
    Sequences~$\{\bx_i^{(t)}\}$, $\{\bz_n^{(t)}\}$ generated by FedBCD satisfy
    \begin{align*}
        \frac{1}{T}\sum_{t=0}^{T-1}\mathbb{E}\left[\left\langle\nabla_{\bx_i}f_i(\bx_i^{(t+1)},\bar{\bz}^{(t)}),\hat{\bx}_i-\bx_i^{(t+1)}\right\rangle\right] &\geq -\frac{A_1}{T^{1/4}}-\frac{A_2}{R^{1/4}},~\forall \hat{\bx}\in\mathcal{X},~i\in\mathcal{Q}_n,n\in\mathcal{V},\\
        \frac{1}{T}\sum_{t=0}^{T-1}\mathbb{E}\left[\left\|\sum_{n\in\mathcal{V}}\sum_{i\in\mathcal{Q}_n}\gamma_i(\bz^{(t)}-\bx_i^{(t+1)})\right\|^2\right]  &\leq \frac{B_1}{\sqrt{T}}+\frac{B_2}{\sqrt{R}},\\
        \frac{1}{T}\sum_{t=0}^{T-1}\|\bar{\bz}^{(t)}-\bz_n^{(t)}\| &\leq\frac{C}{\sqrt{T}},~\forall n\in\mathcal{V},
    \end{align*}
    where $A_1,A_2,B_1,B_2,C$ are positive constants (determined by parameters in Assumption~\ref{Asm:subset}-\ref{asm:batch_size}) whose accurate forms are in the supplementary document, and $\bar{\bz}^{(t)}=\frac{1}{|\mathcal{V}|}\sum_{n\in\mathcal{V}}\bz_n^{(t)}$.
    \item
    If we use the exact gradients during updates, the above results hold with all terms related to the batch size $R$ removed.
\end{enumerate}
\end{Theorem}
The proof of Theorem~\ref{thm} is in Appendix~\ref{sec:appendx}.
We see that the batch size $R$ and communication round $T$ determine the convergence rate, which is the case in projection-based stochastic algorithms \cite{ghadimi2016mini}.
Note that on edge devices the data size is not large and thus exact gradients are possibly computable.
For this case, our result reveals a sub-linear convergence rate of $\mathcal{O}(1/\sqrt{T})$, which is consistent with the existing decentralized non-convex algorithms~\cite{li2018federated,pmlr-v97-assran19a}.

\begin{Remark}
    In our analysis, we discuss the convergence of both the Sync-cloud and Async-cloud settings in a unified way.
    If we only consider the Sync-cloud setting, FedBCD will become a special case of a centralized computation scheme~\cite{wu2020hybrid}.
    By modifying the assumptions to make FedBCD fit this computation scheme, we will be able to obtain a $\mathcal{O}(1/T)$ sub-linear convergence rate, which is standard for centralized non-convex algorithms.
    On the other hand, it should be pointed out that the Async-cloud setting of FedBCD is not seen in the literature, and requires careful and non-trivial analysis.
\end{Remark}

\section{Numerical Experiments}
\label{sec:exp}
We train two models on PyTorch: a three-layer neural network for the classification of~the MNIST dataset~\cite{lecun1998gradient} and a deeper ResNet-20 model~\cite{he2016deep} for the CIFAR-10 dataset~\cite{krizhevsky2009learning}.
We assume that~there are 10 cloud servers, each connected to 10 edge devices.
For the MNIST digit recognition task, training data are randomly distributed with each device receiving 600 data pairs;
for the CIFAR-10 classification task, each device has 500 data pairs.
To emulate the data heterogeneity, we restrict the ``diversity'' of training data;
e.g., ``diversity'' being $3$ means that each edge device owns data of three labels.
We also use a small $|\mathcal{Q}_n^{(t)}|$ to emulate occurrences of intermittent communication.
The other~settings~are:
the edge-device learning rate is $\eta_x=0.005$, the momentum weight $\zeta$ is 0.9,
the edge-device update epoch is sampled from $[1,5]$ every round,
$\mathcal{X}$ is an elementwise $[-2,2]$-box constraint,
the batch size is $R=32$;
for FedBCD/FedBCD-I, the cloud-server learning rate is $\eta_{z}=0.5$;
for FedBCD, the penalty parameter is $\gamma_i=1$;
for FedBCD-I, we~use~$\gamma_i=0.2$,~$|\tilde{\mathcal{Q}}_n^{(t)}|=8$, and edge devices suspend local training after finishing 4 offline rounds;
for FedProx, ASPG is used in edge-device update, and the penalty parameter is $5$.
{The above parameters selection was by trial and error.}

\subsection{Model Deviation vs. Model Consensus}
\begin{figure}[!h]
\centering
    $\bullet$ (upper row) the average {\bf personalized} performance, (lower row) the {\bf global} performance \hfill~\\
    \begin{minipage}[b]{.24\linewidth}
        \centering
        \includegraphics[width=\linewidth]{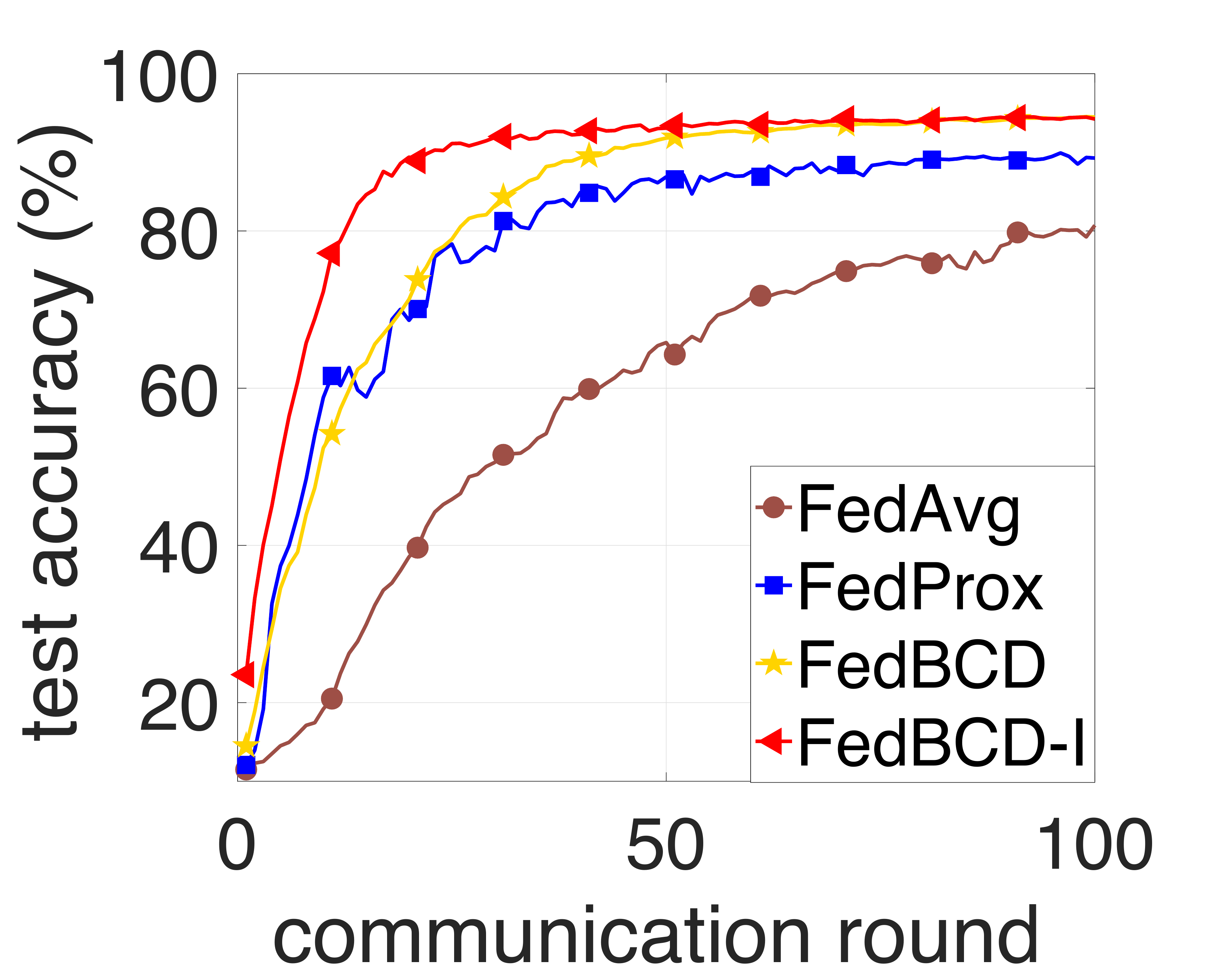}
    \end{minipage}
    \begin{minipage}[b]{.24\linewidth}
        \centering
        \includegraphics[width=\linewidth]{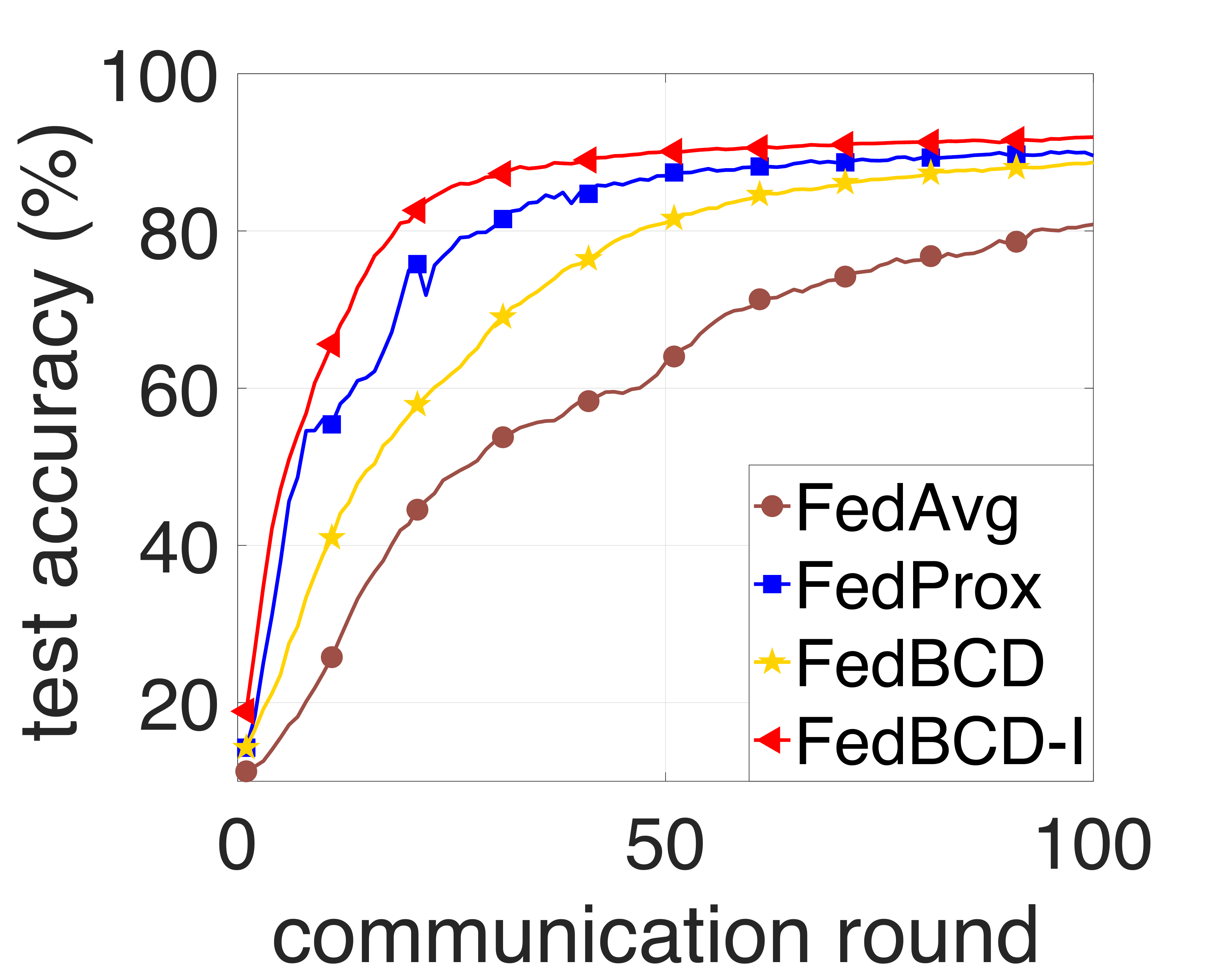}
    \end{minipage}
    \begin{minipage}[b]{.24\linewidth}
        \centering
        \includegraphics[width=\linewidth]{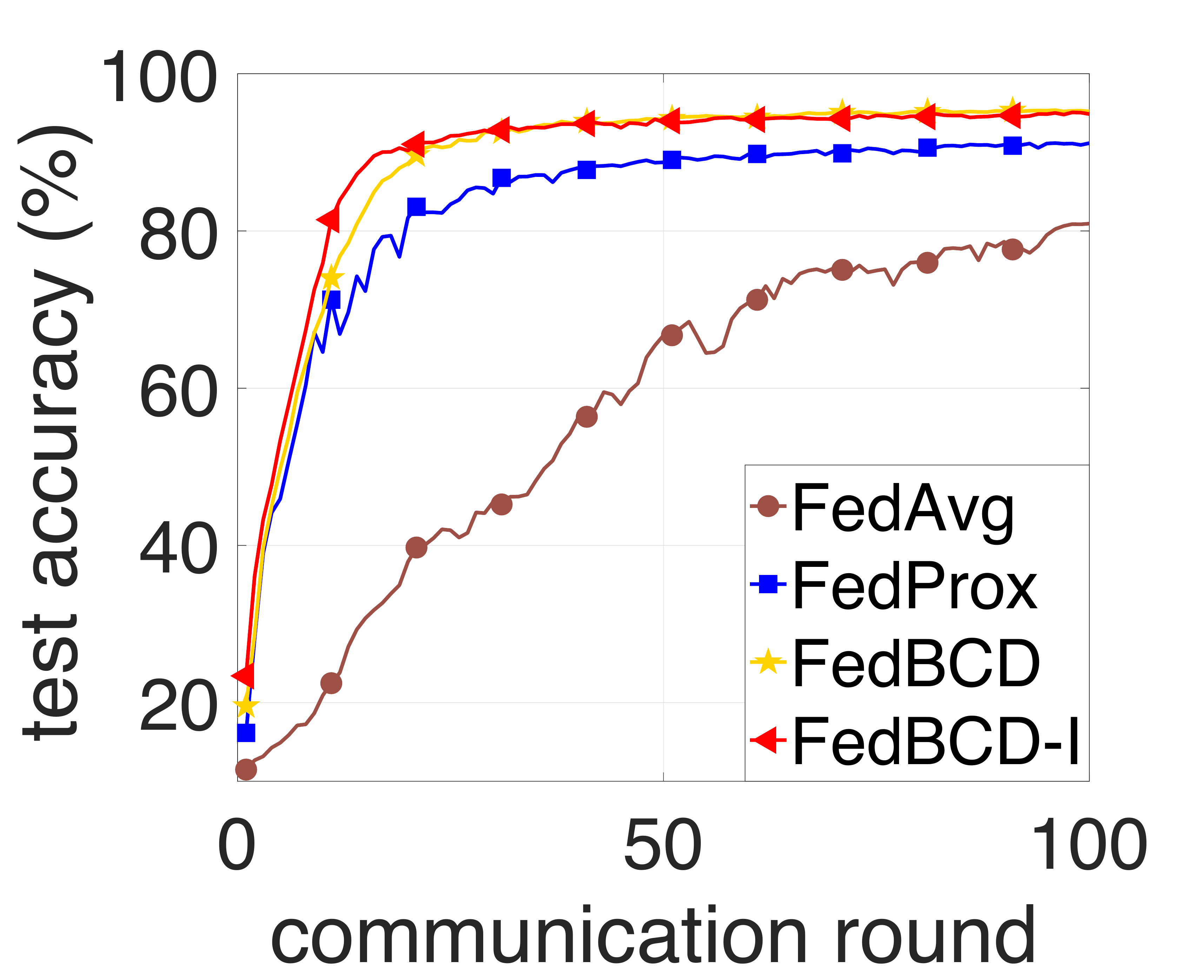}
    \end{minipage}
    \begin{minipage}[b]{.24\linewidth}
        \centering
        \includegraphics[width=\linewidth]{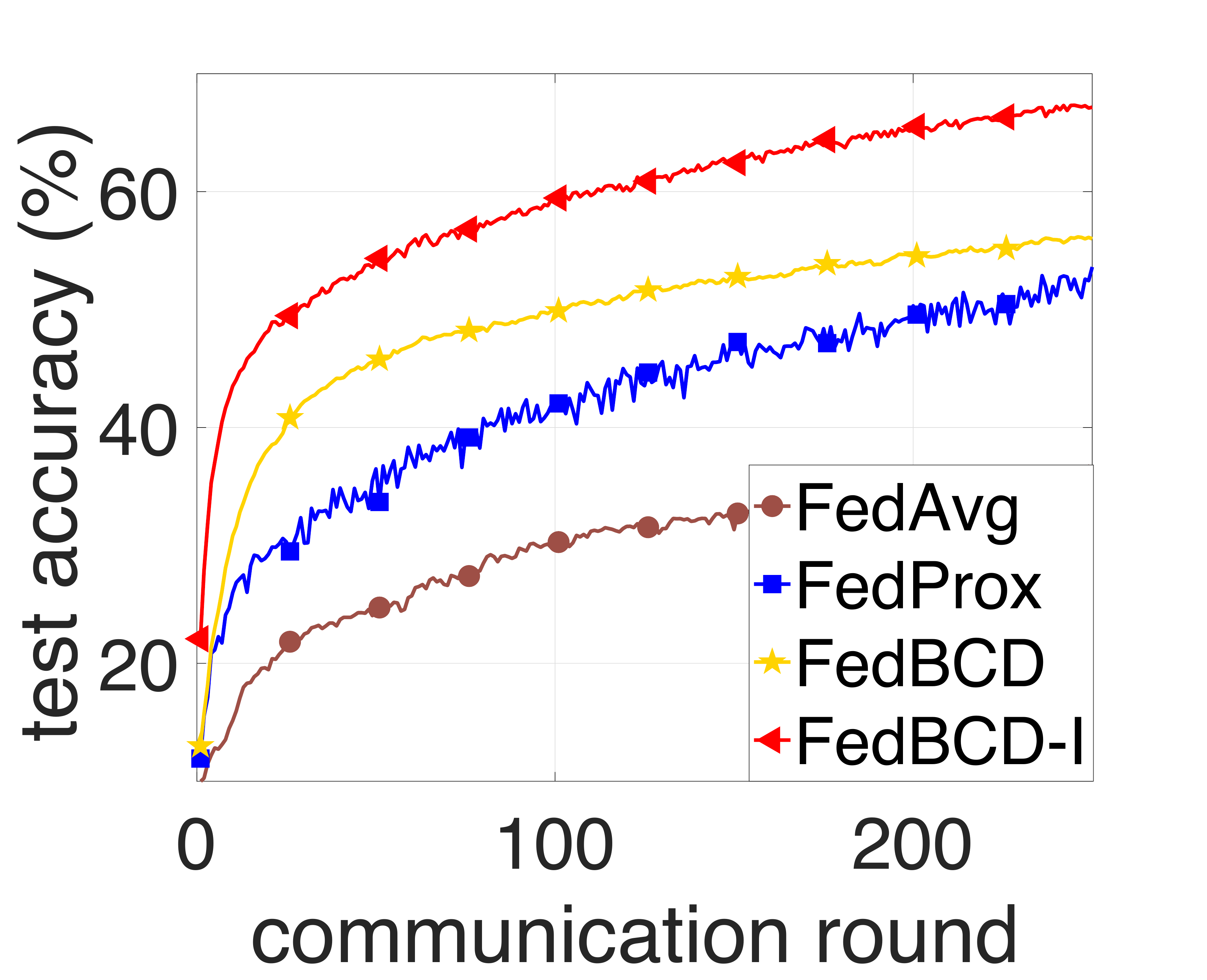}
    \end{minipage}\\
    \begin{minipage}[b]{.24\linewidth}
        \centering
        \includegraphics[width=\linewidth]{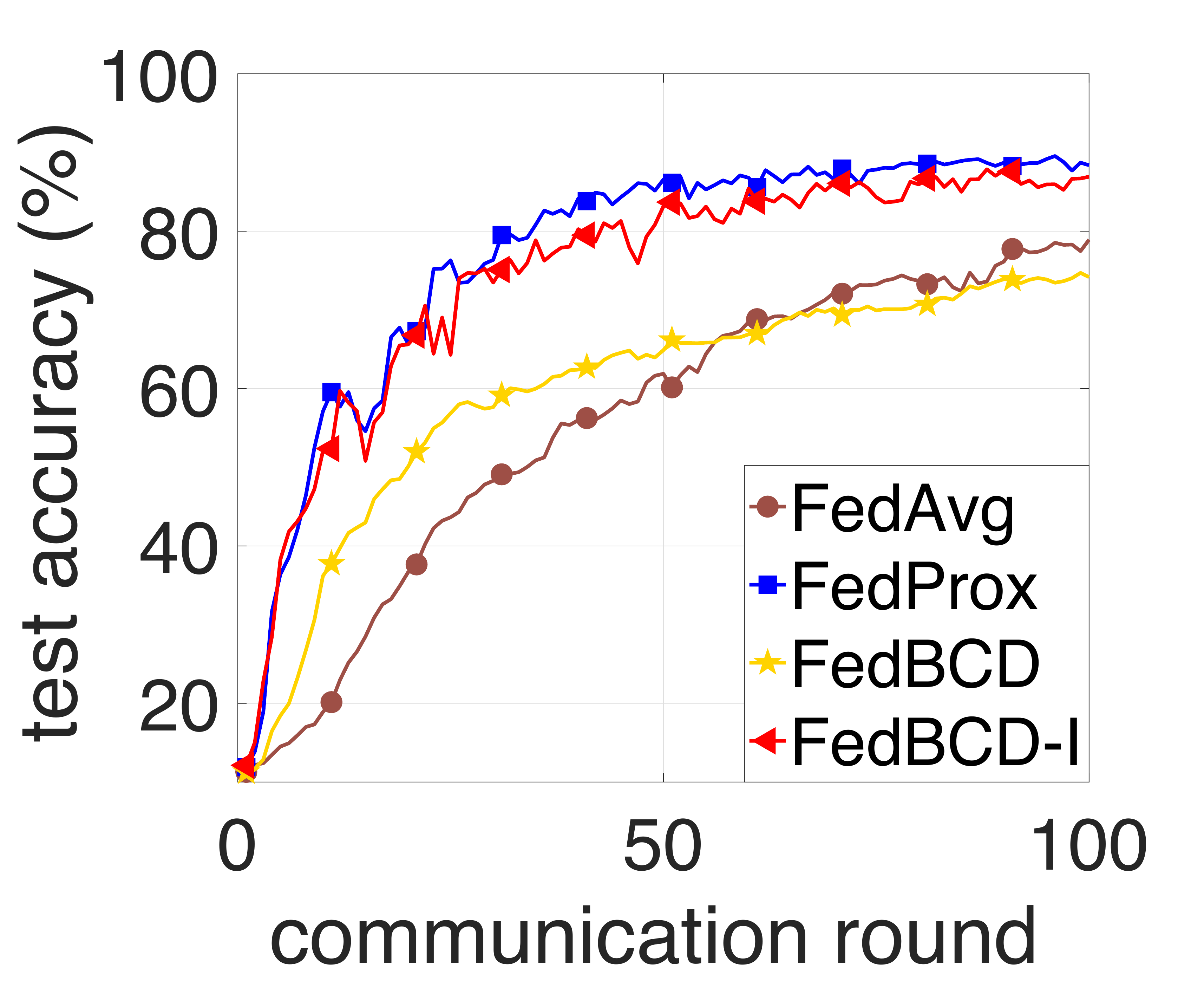}\\
        (a)
    \end{minipage}
    \begin{minipage}[b]{.24\linewidth}
        \centering
        \includegraphics[width=\linewidth]{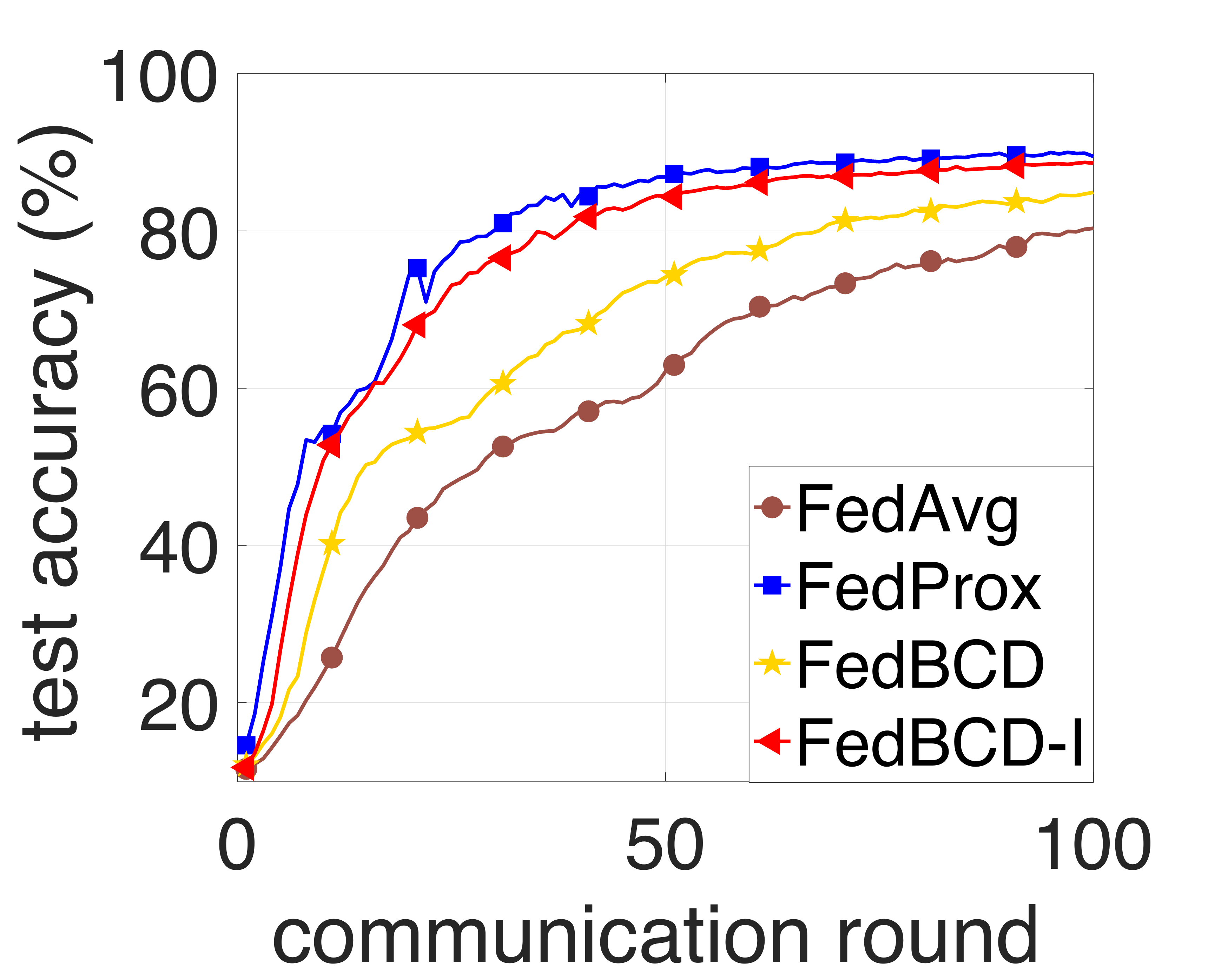}\\
        (b)
    \end{minipage}
    \begin{minipage}[b]{.24\linewidth}
        \centering
        \includegraphics[width=\linewidth]{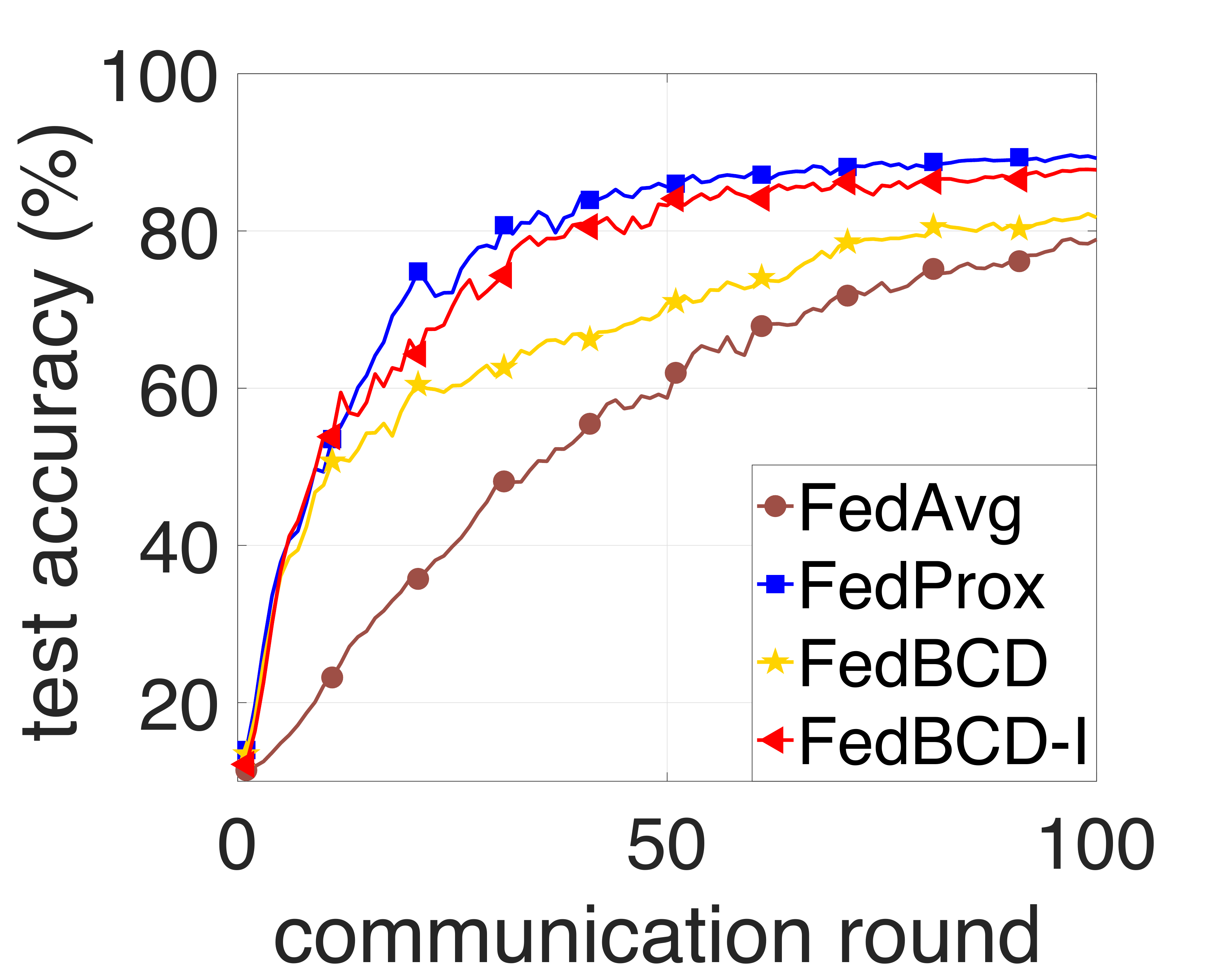}\\
        (c)
    \end{minipage}
    \begin{minipage}[b]{.24\linewidth}
        \centering
        \includegraphics[width=\linewidth]{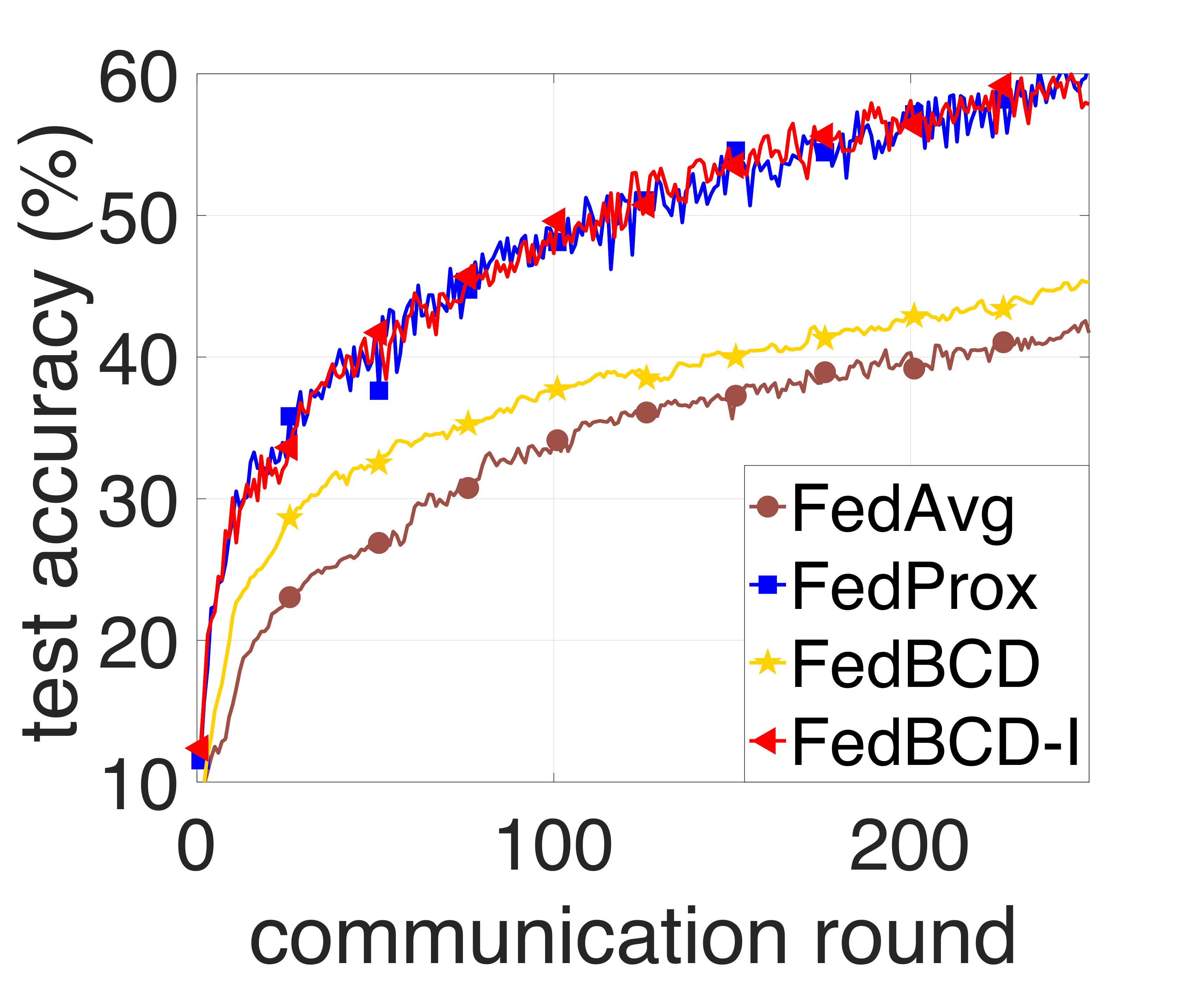}\\
        (d)
    \end{minipage}
    \caption{Test accuracy of models by different algorithms with different diversity and activation. Settings of the experiments (a) MNIST-$|\mathcal{Q}_n^{(t)}|=3$, diversity: 3; (b) MNIST-$|\mathcal{Q}_n^{(t)}|=3$, diversity: 6; (c) MNIST-$|\mathcal{Q}_n^{(t)}|=6$, diversity: 3; (d)~CIFAR-10-$|\mathcal{Q}_n^{(t)}|=5$, diversity: 3.}
    \label{fig.MNIST}
\end{figure}

We compare FedBCD/FedBCD-I (which~allow model deviations) with FedAvg/FedProx~(which enforce model consensus).
Two performance measures are considered:
the personalized performance, where edge devices test local models on test data that have the same diversity as their~training data;
and the global performance, where the cloud tests the global model on the whole test~data set (which has $10,000$ data pairs for both the MNIST and CIFAR-10 tasks).
To be fair, in this experiment all algorithms use the Sync-cloud protocol\footnote{The Sync-cloud protocol for FedAvg/FedProx is as follows: (i) A new round~$t$ starts with the ``response and update'' phase.
In this phase, cloud servers accept update requests~from edge devices, send the global model $\bz^{(t)}$ to these activated edge devices, and wait for them to~perform the local updates.
For each cloud server, this phase ends when it receives $\bx_i^{(t+1)}$'s from a pre-determined number of edge devices (this subset of edge devices is named $\mathcal{Q}_n^{(t)}$).
Then, this cloud server sends the sum of received edge-device models $\sum_{i\in\mathcal{Q}_n^{(t)}}\bx_i^{(t+1)}$ to a coordinator.
(ii)
Once hearing from all cloud servers, this coordinator enters the ``aggregation'' phase,
where it calculates the average of all received edge-device models $\bx_i^{(t+1)}$ to obtain $\bz^{(t+1)}$~and broadcasts it back.
After receiving $\bz^{(t+1)}$, the cloud servers become available again for the next round of update.}, and the two types of performance are recorded every round (i.e. the latency per round is not considered).
It can be seen in Fig.~\ref{fig.MNIST} that, the personalized and global performance of FedAvg/FedProx is almost the same.
This makes sense, since therein the same model is maintained everywhere.
In comparison, FedBCD/FedBCD-I, by allowing local models to deviate from the global one, provide better personalized performance.
We also see that FedBCD suffers from slow update progress when the $|\mathcal{Q}_n^{(t)}|$ is low and is only faster than FedAvg (this is not surprising, see Section~\ref{sec:variant}).
In contrast, its intuitive variant FedBCD-I turns out to be more competitive;
particularly, we see that in the more challenging CIFAR-10 task, FedBCD-I has global performance comparable to FedProx, while giving much better personalized performance.

\subsection{Sync-Cloud  vs. Async-Cloud}~~
Following the latency analysis in Section~\ref{sec:protocol}, we demonstrate the possible efficiency gain of the Async-cloud update.
To emulate the distributed nature of the experiment, we implemented each edge device, cloud server and the coordinator as separate Python programs.
For each simulation we spin up a separate process for the coordinator, each cloud server and each edge device, and the processes communicate over plain TCP sockets.
Simulations were run on four Dell R920 60 core servers with Intel Xeon E7-4870 v2 CPUs and 256 GB of memory each, running Debian Linux, with the servers physically located in the same rack and interconnected using a 10 Gbit/s Ethernet.
The 111 processors (100 edge devices+10 cloud servers+1 coordinator) were distributed across the servers to ensure that no server was ever at full utilization (all cores occupied simultaneously) at any point in time, which would create artificial delays in the solving time.
No delay is added to the TCP communication, as we assume the latency between cloud servers to be insignificant compared with the edge computational time, and a reasonable bandwidth between (transcontinental) datacenters is on the order of 10 Gbit/s or even faster.
Exponentially distributed delays are added to the edge processes to simulate variations in computational capacity across the edge devices and the edge to cloud latency\footnote{Specifically, we generate numerically the random latency $\tau_n^{(t)}$ for the ``respond and update'' phase (green phases in Figure \ref{fig:protocol}) and assume that server aggregation computation (non-green phases in Figure \ref{fig:protocol}) is negligible, i.e. the overall runtime is dominated by $\tau_n^{(t)}$.
To be clear, at the round $t$, each edge device simulates an inter-arrival time $\tau_{i,{\sf arrival}}^{(t)}$ for its update request to the cloud, using an exponential distribution variable (with mean $2$).
The cloud server will then build the activated edge device set $\mathcal{Q}_n^{(t)}$ including edge devices that have the shortest arrival time ($|\mathcal{Q}_n^{(t)}|$ is fixed for each experiment to 3, 5 and 6).
Then, the selected edge device $i$ will implement local training using a random number of epochs $K_i^{(t)}$ with range $[1, 5]$.
The processing time of each epoch $\tau_{i,{\sf process}}^{(t)}$ is simulated by another exponential distribution with mean $1$.
Then, the total latency of an activated edge device $i$ is $\tau_i^{(t)}=\tau_{i,{\sf arrival}}^{(t)}+K_i^{(t)}\tau_{i,{\sf process}}^{(t)}$.
Finally, the latency of cloud server $n$ is $\tau_n^{(t)}=\max_{i\in\mathcal{Q}_n^{(t)}}\tau_i^{(t)}$.}.
\begin{figure}[hbt]
\centering
    $\bullet$ the {\bf global} performance:\hfill~\\
    \begin{minipage}[b]{.24\linewidth}
        \centering
        \includegraphics[width=\linewidth]{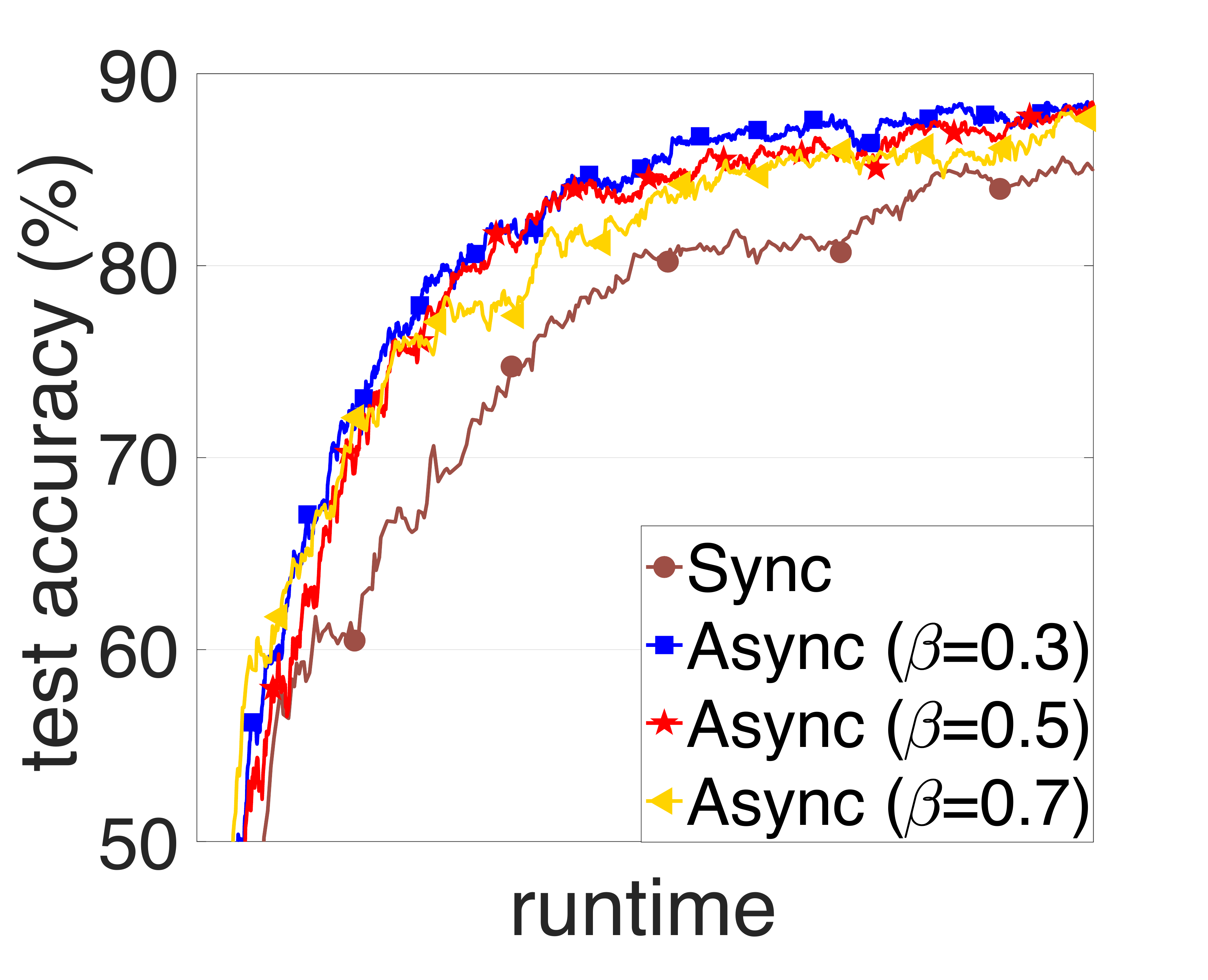}
        (a) FedBCD
    \end{minipage}
    \begin{minipage}[b]{.24\linewidth}
        \centering
        \includegraphics[width=\linewidth]{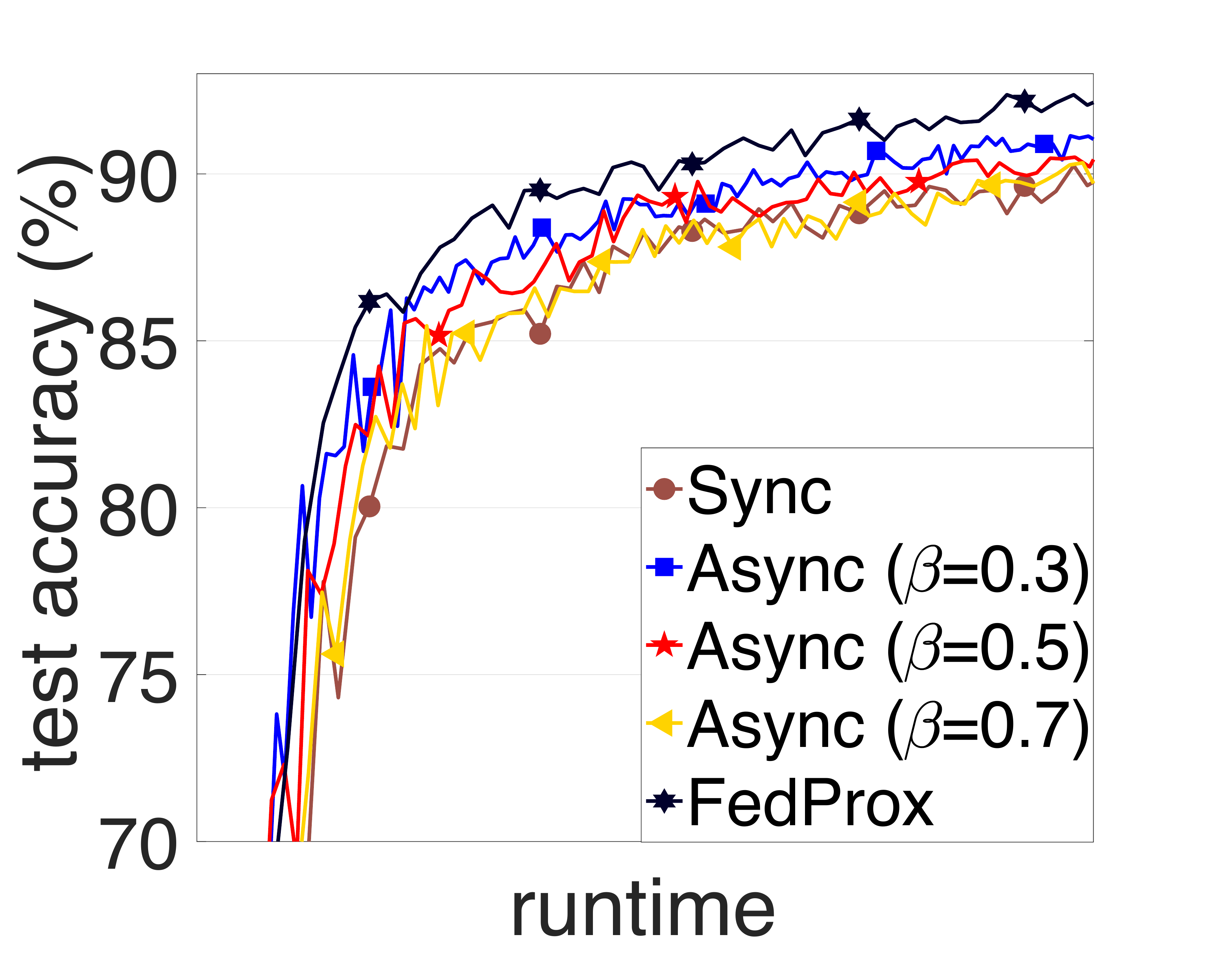}
        (b) FedBCD-I
    \end{minipage}
    \begin{minipage}[b]{.24\linewidth}
        \centering
        \includegraphics[width=\linewidth]{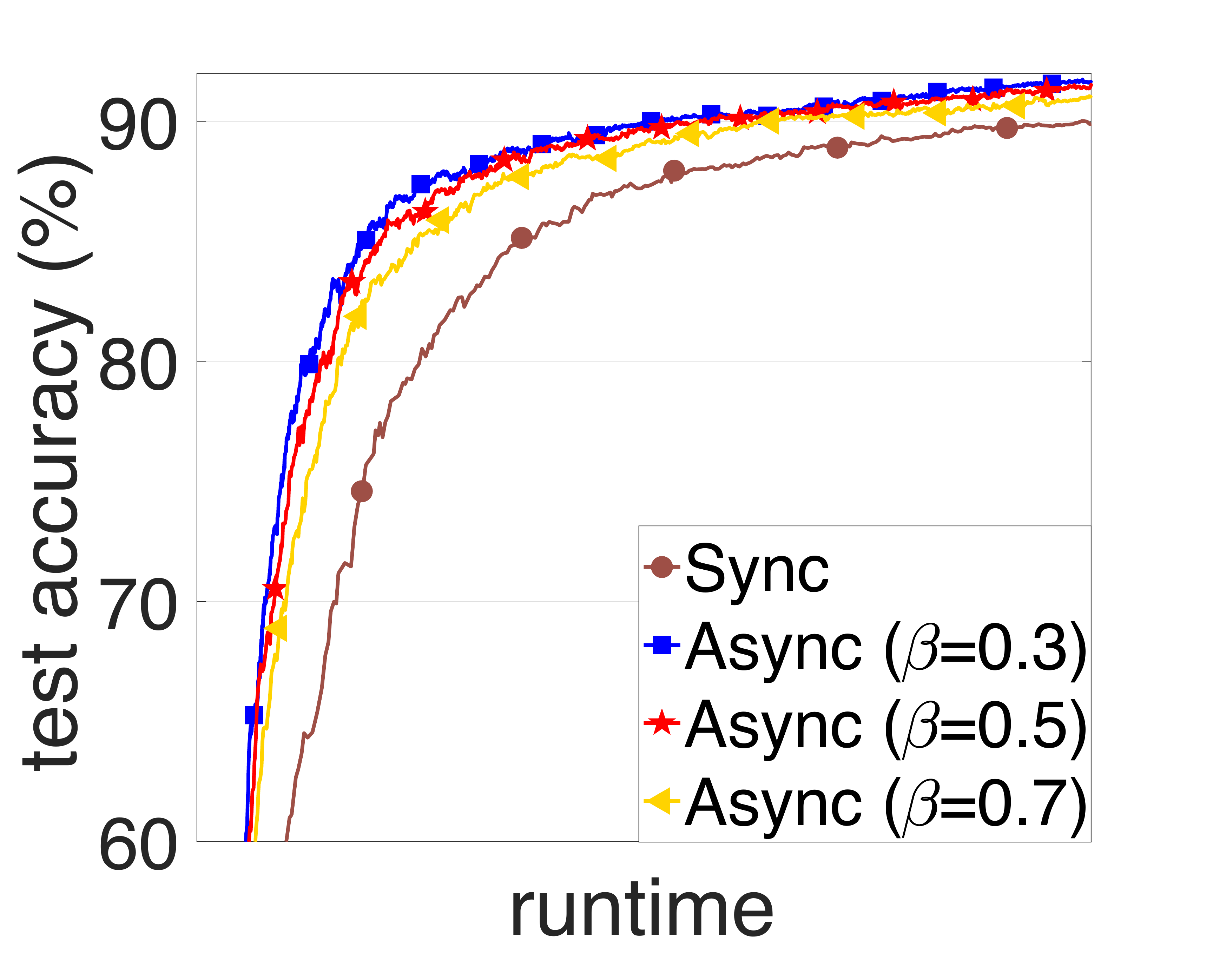}
        (c) FedBCD
    \end{minipage}
    \begin{minipage}[b]{.24\linewidth}
        \centering
        \includegraphics[width=\linewidth]{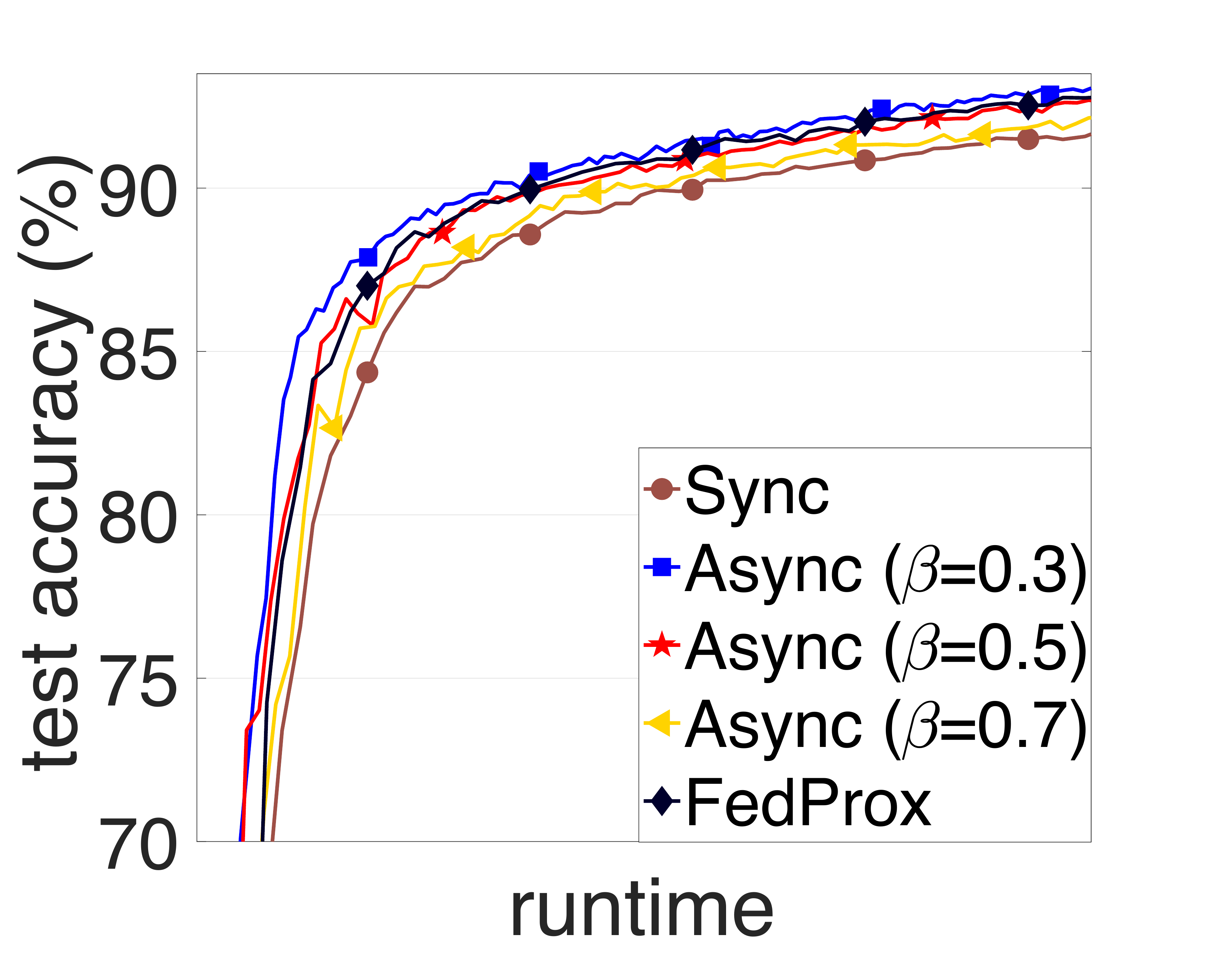}
        (d) FedBCD-I
    \end{minipage}
    \caption{Efficiency of the algorithms, on MNIST with $|\mathcal{Q}_n^{(t)}|=3$ and ``diversity ''(a-b) 3, (c-d) 6}.
    \label{eq:fig_runtime}
\end{figure}

The results are shown in Fig.~4.
We first notice that, the higher ``diversity'' leads to a smoother improvement of the global performance.
This is reasonable, since a more ``biased'' local model is expected to bring more ``noise'' to the global model training.
Besides, we can see that the Async-cloud update is clearly more efficient.
As mentioned before, based on the size of $\beta$, there is a trade-off between the duration and the update progress of each round.
In this experiment, we see that $\beta=0.3$ or $0.5$ seems to strike a good balance and leads to appealing performance.
We also compare FedBCD-I with FedProx.
Note that FedProx focuses on improving the global performance, and is generally faster than our methods for the global performance.
However, by using the asynchronous update, our scheme could outperform FedProx even for the global performance in terms of runtime;
see Fig.~\ref{eq:fig_runtime}(d).

\section{Conclusion}
This work has two main contributions:
first, we recognize that machine learning models on edge devices reflect the user habits, and can thus be different;
second, we spotlight that the cloud is a cluster of powerful servers, and their intra-communication can be exploited for more efficient updates.
We study these two aspects and provide a solution that is proven promising theoretically and empirically.

\section*{Acknowledgment}
This work was supported by the Army Research Office, USA, under Project ID ARO \#W911NF-20-1-0153.
The work by Ruiyuan Wu and Wing-Kin Ma was supported by a General Research Fund (GRF) of the Research Grant Council (RGC), Hong Kong, under Project ID CUHK 14208819.

\section*{Appendix}
\appendix
\begin{appendix}
\section{An alternative communication protocol for FedBCD}
\label{appendix:1}
As mentioned in Section~\ref{Sec:Anal}, the Async-cloud protocol introduced in Section~\ref{sec:protocol} violates our assumptions.
To explain, the theoretical analysis requires all $\eta_{z_n}^{(t)}$ to be the same at each round, while the protocol forces some of them to be $0$ if they are not the fastest $B$ cloud servers.
Here, we provide a slightly more complicated version that strictly follows the assumption.
\paragraph*{\it $\bullet$ Async-cloud protocol (rigorous).}
(i) At round $t$, the cloud server that finished the ``response and update'' phase directly sends its model $\bz_n^{(t)}$ to the coordinator (only cloud-server models at round $t$ will be accepted).
(ii) The coordinator waits until it hears from the first $B$ cloud servers $\mathcal{V}^{(t)}$ [in Fig.~\ref{fig:protocol}(b) it is $|\mathcal{V}^{(t)}|=B=2$], and enters ``aggregation'' phase I.
(iii) Then, the coordinator calculates $\bw^{(t)}=\sum_{n\in\mathcal{V}^{(t)}}\bz_n^{(t)}/B$ and sends $\bw^{(t)}$ to cloud servers in $\mathcal{V}^{(t)}$.
It will also send an ``isolation'' signal to cloud servers not in $\mathcal{V}^{(t)}$.
(iv) In ``aggregation'' phase II, cloud servers in $\mathcal{V}^{(t)}$~use received $\bw^{(t)}$ as $\bw_n^{(t)}$ and update $\bz_n^{(t+1)}$ by~\eqref{eq:DGD};
the isolated cloud servers will wait until their ``edge-device update'' phase ends, and then use $\bw_n^{(t)}=\bz_n^{(t)}$ in update~\eqref{eq:DGD} to obtain $\bz_n^{(t+1)}$.

\begin{figure}[h]
    \centering
    \includegraphics[width=\linewidth]{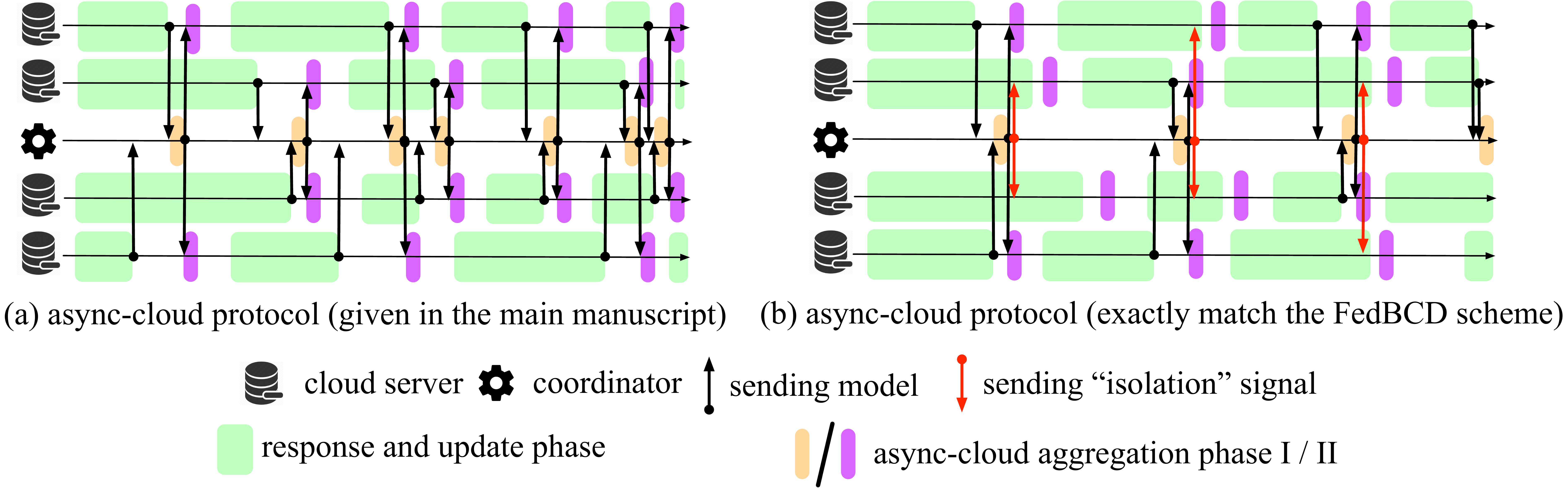}
    \caption{The two Async-cloud protocols.}
    \label{fig:protocol_async}
\end{figure}

The comparison of the two Async-cloud protocols are illustrated in Fig.~\ref{fig:protocol_async}.
We can see that, in the ``rigorous'' Async-cloud protocol, the coordinator will send ``isolation'' signal to cloud servers that are not in $\mathcal{V}^{(t)}$ and ask them to carry out an ``isolated'' update, rather than simply put $\bz_n^{(t+1)}=\bz_n^{(t)}$.
In this way, the same $\eta_{z_n}^{(t)}$ is used on all cloud servers.
In practice, we figure out that the ``isolated'' update makes the protocol more complicated to implement.
Moreover, we notice that these protocols give rise to similar performance.
In view of this, we choose to present the more ``asynchronous'' but less ``rigorous'' protocol in the main manuscript.

\section{Proof of Theorem~\ref{thm}}
\label{sec:appendx}
For notational simplicity, we will fix $K_x=K_z=1$ and omit the epoch index $k$ (since only one epoch of update is carried out in each round under this setting) in FedBCD.
We should emphasize that the subsequent proof can be easily modified to cover the general cases.
Moreover, since the Sync-cloud setting can be regarded as a special setting of Async-cloud setting (see the discussion below Algorithm~\ref{alg:FedBCD_full}), we will only focus on the Async-cloud setting.
Some other notations are listed below.

{\bf Notation.}
$D_{\mathcal{X}}=\max_{\bx\in\mathcal{X},\by\in\mathcal{X}}\|\bx-\by\|$;
$D_{g_i}=\max_{\bx\in\mathcal{X}}\|\nabla g_i(\bx)\|$;
$\gamma=\max_{i\in\mathcal{Q}_n,n\in\mathcal{V}}\gamma_i$;
$\bar{\bz}^{(t)}=(\sum_{n\in\mathcal{V}}\bz_n^{(t)})/|\mathcal{V}|$;
$\bar{\bz}^{(t)}=(\sum_{n\in\mathcal{V}}\bz_n^{(t)})/|\mathcal{V}|$;
let $\bA^{(t)}$ be a consensus matrix with its $(n,m)$th entry being $a_{n,m}^{(t)}$,
and define $\bm\Phi(t,s) = \bA^{(t)}\bA^{(t-1)}\cdots\bA^{(s+1)}$;
by ``for any $i, n, t$'' we mean for any $i\in\mathcal{Q}_n,n\in\mathcal{V}$, $t=0,1,\ldots,T-1$.\\

\subsection{Some Preliminary Facts}
To begin with, recall that we are interested in tackling the following problem:
\begin{equation}
\label{eq:prob_a}
\begin{aligned}
    \min_{\substack{\{\bx_i\}_{i\in\mathcal{Q}_n,n\in\mathcal{V}}\\\{\bz_n\}_{n\in\mathcal{V}}}}~&\sum_{n\in\mathcal{V}}\sum_{i\in\mathcal{Q}_n}f_i(\bx_i,z_n)\\
    {\rm s.t.}\qquad&\bx_i\in\mathcal{X},\forall i\in\mathcal{Q}_n,n\in\mathcal{V},\\
    &\bz_n=\bz_m,~\forall(n,m)\in\mathcal{E},
\end{aligned}
\end{equation}
where $f_i(\bx_i,\bz_n) : = g_i(\bx_i)+\frac{\gamma_i}{2}\|\bx_i-\bz_n\|^2$.

Denote the mini-batch stochastic gradient in ASPG (line~\ref{alg:apg_1} of Algorithm~\ref{alg:FedBCD_full}) as $\nabla_{\bx_i}\tilde{f}_i(\bx_{i,\sf ex}^{(t)},\bz_n^{(t)})$;
i.e.,
\[
    \nabla_{\bx_i}\tilde{f}_i(\bx_{i,\sf ex}^{(t)},\bz_n^{(t)}) = \frac{1}{R}\sum_{r=1}^R\nabla h(\bx_{i,\sf ex}^{(t)};\bxi_{i,r}^{(t)})+\gamma_i(\bx_{i,\rm ex}^{(t)}-\bz_n^{(t)}).
\]
It can also be written as
\begin{equation}
\label{appendx:2}
    \nabla_{\bx_i}\tilde{f}_i(\bx_{i,\sf ex}^{(t)},\bz_n^{(t)}) = \nabla_{\bx_i} f_i(\bx_{i,\sf ex}^{(t)},\bz_n^{(t)})+\bm\delta_i^{(t)},
\end{equation}
where the gradient estimation error is $\bm\delta_i^{(t)} = \frac{1}{R}\sum_{r=1}^R\nabla h(\bx_{i,\rm ex}^{(t)};\bxi_{i,r}^{(t)}) - \nabla g_i(\bx_{i,\rm ex}^{(t)})$.

We first have the following facts.
\begin{Fact}
\label{fact:cvx_comb}
    Suppose that Assumptions~\ref{Asm:function}-\ref{asm:stepsize} hold.
    For any $n$, the sequences $\{\bw_n^{(t)}\}$ and $\{\bz_n^{(t+1)}\}$ generated by Algorithm~\ref{alg:FedBCD_full} always lie on $\mathcal{X}$.
\end{Fact}
{\it Proof:}
Under Assumption \ref{asm:stepsize}, the update rule of $\bz_n^{(t+1)}$ in Algorithm~\ref{alg:FedBCD_full} is always a convex combination of sequences $\bw_n^{(t)}$ and $\{\bx_i^{(t+1)}\}_{i\in\mathcal{Q}_n}$.
Since $\bx_i^{(t+1)}\in\mathcal{X}$ and $\bz_n$ is initialized such that $\bz_n^{(0)}\in\mathcal{X}$, given Assumption~\ref{Asm:5} and the convex combination update of $\bz_n^{(t+1)}$, it is not hard to see that $\bw_n^{(t)}$ and $\bz_n^{(t+1)}$ are guaranteed to lie on $\mathcal{X}$.
$\hfill\blacksquare$

\begin{Fact}
\label{fact:stochastic error}
Suppose that Assumption~\ref{asm:batch_size} holds.
The gradient estimation error defined in~\eqref{appendx:2} satisfies that $\mathbb{E}[\bm\delta_i^{(t)}]=\bm0$ and $\mathbb{E}[\|\bm\delta_i^{(t)}\|^2]=\frac{\sigma^2}{R}$, for all $i,t$.
\end{Fact}
{\it Proof:}
Invoking Assumption~\ref{asm:batch_size}, it can be checked that
\begin{align*}
    \mathbb{E}[\bm\delta_i^{(t)}]=&\frac{1}{R}\sum_{r=1}^{R}\mathbb{E}[\nabla h(\bx_{i,\sf ex}^{(t)};\bxi_{i,r}^{(t)})-\nabla g_i(\bx_{i,\sf ex}^{(t)})] = \bm0,\\
    \mathbb{E}[\|\bm\delta_i^{(t)}\|^2]
    =&\frac{1}{R^2}\left(\mathbb{E}[\|\nabla h(\bx_{i,\sf ex}^{(t)};\bxi_{i,1}^{(t)})-\nabla g_i(\bx_{i,\sf ex}^{(t)})\|^2]+\mathbb{E}\left[\left\|\sum_{r=2}^{R}(\nabla h(\bx_{i,\sf ex}^{(t)};\bxi_{i,r}^{(t)})-\nabla g_i(\bx_{i,\sf ex}^{(t)}))\right\|^2\right]\right.\\
    \quad&\left.+2\mathbb{E}\left\langle\nabla h(\bx_{i,\sf ex}^{(t)};\bxi_{i,1}^{(t)})-\nabla g_i(\bx_{i,\sf ex}^{(t)}),\sum_{r=2}^{R}(\nabla h(\bx_{i,\sf ex}^{(t)};\bxi_{i,r}^{(t)})-\nabla g_i(\bx_{i,\sf ex}^{(t)}))\right\rangle\right)\\
    \leq&\frac{1}{R^2}\left(\sigma^2+\mathbb{E}\left[\left\|\sum_{r=2}^{R}(\nabla h(\bx_{i,\sf ex}^{(t)};\bxi_{i,r}^{(t)})-\nabla g_i(\bx_{i,\sf ex}^{(t)}))\right\|^2\right]\right)\\
    \leq& \ldots
    \leq \frac{\sigma^2}{R},
\end{align*}
where in the first inequality we use $\mathbb{E}[\nabla h(\bx_{i,\sf ex}^{(t)};\bxi_{i,1}^{(t)})]-\nabla g_i(\bx_{i,\sf ex}^{(t)})=\bm0$.
The proof is complete.$\hfill\blacksquare$

\begin{Fact}
\label{fact:lips}
    Suppose that Assumption~\ref{Asm:function} holds.
    For any $i,n$, it holds that
    \begin{enumerate}[leftmargin=0cm,itemindent=.5cm,labelwidth=\itemindent, nolistsep,labelsep=0cm,align=left,label=(\roman*)]
      \item
      $f_i$ is $(L_i+\gamma_i)$-Lipschitz smooth w.r.t. $\bx_i$ on $\tilde{\mathcal{X}}$, $\gamma_i$-Lipschitz smooth w.r.t. $\bz_n$.
      \item
      $f_i$ is $\rho_i$-weakly convex w.r.t. $\bx_i$ on $\tilde{\mathcal{X}}$;
      $f_i(\bx_i,\bz_n)+\frac{\rho_i}{2}\|\bx_i\|^2$ is convex w.r.t. $\bx_i$ on $\tilde{\mathcal{X}}$.
      We have that $\rho_i\leq L_i+\gamma_i$;
      Moreover, if $g_i$ is convex, we have $\rho_i=0$.
    \end{enumerate}
\end{Fact}
{\it Proof:
The Lipschitz smoothness is obvious given Assumption~\ref{Asm:function}
, and the weak convexity of $f_i$ comes with the Lipschitz smoothness~\cite{dong2014proximal}.
$\hfill\blacksquare$
}

\subsection{One-Step Progress of Both Edge-Device and Cloud-Server Update}
In the sequel, we will demonstrate that some sort of ``sufficient decrease'' of the objective functions is achieved for both $\bx_i$-blocks and $\{\bz_n\}_{n\in\mathcal{V}}$-block updates.
\paragraph*{$\bx_i$-block updates.}
We first characterize the progress made in one step of the ASPG update for edge device that is activated; i.e., lines~\ref{alg:apg_3}-\ref{alg:apg_2} in Algorithm~\ref{alg:FedBCD_full}.
To simplify the notations, we will use only in this subsection that
\begin{align*}
    &\bx_i^+=\bx_i^{(t+1)},~\bx_i=\bx_i^{(t)},~\bx_i^-=\bx_i^{(t^-)},~\bx_{i,\sf ex}=\bx_{i,\sf ex}^{(t)},\bz_n=\bz_n^{(t)},~\bm\delta_i=\bm\delta_i^{(t)},
\end{align*}
for $i\in\mathcal{Q}_n^{(t)}$, where $\bx_i^{(t^-)}$ denotes the last active round when edge device $i$ carries out update.
Using the above notations and~\eqref{appendx:2}, the ASPG update can be expressed as
\begin{subequations}
\label{appendx:1}
\begin{align}
\label{eq:up_x_1}
    \bx_{i,\sf ex} &= \bx_i+\zeta (\bx_i-\bx_i^-),\\
\label{eq:up_x_2}
    \bx_i^+ &= \Pi_{\mathcal{X}}\left(\bx_{i,\sf ex}-\eta_x\nabla_{\bx_i}\tilde{f}_i(\bx_{i,\sf ex},\bz_n)\right).
\end{align}
\end{subequations}
Given~\eqref{appendx:1}, we have the following fact.
\begin{Fact}
\label{fact:1}
    The update $\bx_i^+$ generated by~\eqref{appendx:1} satisfies that
    \[
        \bx_i^+ = \min_{\hat{\bx}_i\in\mathcal{X}}\left\langle\nabla_{\bx_i}\tilde{f}_i(\bx_{i,\sf ex},\bz_n),\hat{\bx}_i\right\rangle+\frac{1}{2\eta_x}\left\|\hat{\bx}_i-\bx_{i,\sf ex}\right\|^2.
    \]
    Also, the following first-order optimality condition holds
    \[
        \left\langle\nabla_{\bx_i}\tilde{f}_i(\bx_{i,\sf ex},\bz_n)+\frac{1}{\eta_x}(\bx_i^+-\bx_{i,\sf ex}),\hat{\bx}_i-\bx_i^+\right\rangle\geq 0,~\forall\hat{\bx}_i\in\mathcal{X}.
    \]
\end{Fact}

Given Fact~\ref{fact:1}, we have the following lemma.
\begin{Lemma}\label{Lemma:sufficient_decrease_xi}
Suppose that Assumptions~\ref{Asm:function}-\ref{Asm:5} hold.
Then, for any edge device $i$ that is activated, the update $\bx_i^+$ generated by~\eqref{appendx:1} satisfies that
\begin{align*}
    \frac{1}{2\eta_x}\|\bx_i^+-\bx_i\|^2
    \leq& f_i(\bx_i,\bar{\bz})-f_i(\bx_i^+,\bar{\bz})+\zeta ^2\left(\frac{1}{2\eta_x}+\frac{\rho_i}{2}\right)\|\bx_i-\bx_i^-\|^2+\\
    &(\zeta D_{\mathcal{X}}+\eta_x(D_{g_i}+\gamma_i D_{\mathcal{X}}+\|\bm\delta_i\|))(\|\bm\delta_i\|+\gamma_i\|\bar{\bz}-\bz_n\|).
\end{align*}
\end{Lemma}
{\it Proof:}
To begin with, from the Lipschitz smoothness and weakly convexity of $f_i$ w.r.t. $\bx_i$ (in Fact~\ref{fact:lips}), we have
\begin{align*}
    f_i(\bx_i^+,\bz_n)
    &\leq f_i(\bx_{i,\sf ex},\bz_n)+\left\langle\nabla_{\bx_i} f_i(\bx_{i,\sf ex},\bz_n),\bx_i^+-\bx_{i,\sf ex}\right\rangle+\frac{L_i+\gamma_i}{2}\|\bx_i^+-\bx_{i,\sf ex}\|^2, \\
    f_i(\bx_i,\bz_n) &\geq f_i(\bx_{i,\sf ex},\bz_n)+\langle\nabla_{\bx_i} f_i(\bx_{i,\sf ex},\bz_n),\bx_i-\bx_{i,\sf ex}\rangle-\frac{\rho_i}{2}\|\bx_i-\bx_{i,\sf ex}\|^2.
\end{align*}
Combining the above inequalities and $\eta_x\leq\frac{1}{L_i+\gamma_i}$ (see Assumption~\ref{asm:stepsize}(i)), we obtain
\begin{align}
    f_i(\bx_i^+,\bz_n)
    \leq& f_i(\bx_i,\bz_n)+\left\langle\nabla_{\bx_i} f_i(\bx_{i,\sf ex},\bz_n),\bx_i^+-\bx_i\right\rangle+\frac{1}{2\eta_x}\|\bx_i^+-\bx_{i,\sf ex}\|^2+\frac{\rho_i}{2}\|\bx_i-\bx_{i,\sf ex}\|^2 \nonumber\\
    \leq& f_i(\bx_i,\bz_n)+\left\langle\bm\delta_i+\frac{1}{\eta_x}(\bx_i^+-\bx_{i,\sf ex}),\bx_i-\bx_i^+\right\rangle+\frac{1}{2\eta_x}\|\bx_i^+-\bx_{i,\sf ex}\|^2+\frac{\rho_i}{2}\|\bx_i-\bx_{i,\sf ex}\|^2 \nonumber\\
    \leq& f_i(\bx_i,\bz_n)-\frac{1}{2\eta_x}\|\bx_i^+-\bx_i\|^2 + \|\bm\delta_i\|\|\bx_i^+-\bx_i\|+\left(\frac{1}{2\eta_x}+\frac{\rho_i}{2}\right)\|\bx_i-\bx_{i,\sf ex}\|^2 \nonumber\\
    =& f_i(\bx_i,\bz_n)-\frac{1}{2\eta_x}\|\bx_i^+-\bx_i\|^2 + \|\bm\delta_i\|\|\bx_i^+-\bx_i\|+\zeta ^2\left(\frac{1}{2\eta_x}+\frac{\rho_i}{2}\right)\|\bx_i-\bx_i^-\|^2 \nonumber\\
    \leq& f_i(\bx_i,\bz_n)-\frac{1}{2\eta_x}\left\|\bx_i^+-\bx_i\right\|^2 + \|\bm\delta_i\|(\zeta D_{\mathcal{X}}+\eta_x(D_{g_i}+\gamma_i D_{\mathcal{X}}+\|\bm\delta_i\|))\nonumber\\
\label{eq:sufficient_x_1}
    &+\zeta ^2\left(\frac{1}{2\eta_x}+\frac{\rho_i}{2}\right)\|\bx_i-\bx_i^-\|^2,
\end{align}
where the second inequality is due to Fact~\ref{fact:1} and~\eqref{appendx:2},
the third inequality uses that
\[
    \langle\bx_i^+-\bx_{i,\sf ex},\bx_i-\bx_i^+\rangle=\frac{1}{2}\left(\|\bx_i-\bx_{i,\sf ex}\|^2-\|\bx_i^+-\bx_{i,\sf ex}\|^2-\|\bx_i^+-\bx_i\|^2\right),
\]
the equality uses~\eqref{eq:up_x_1}, and
the last inequality is based on
\begin{align}
    \label{eq:diff_x}
    \left\|\bx_i-\bx_i^+\right\|
    =&\left\|\bx_i-\bx_{i,\sf ex}+\bx_{i,\sf ex}-\bx_i^+\right\|\nonumber\\
    \leq&\zeta \|\bx_i-\bx_i^-\|+\eta_x\left\|\nabla g_i(\bx_{i,\sf ex})+\gamma_i(\bx_{i,\sf ex}-\bz_n)+\bm\delta_i\right\|\nonumber\\
    \leq&\zeta D_{\mathcal{X}}+\eta_x(D_{g_i}+\gamma_i D_{\mathcal{X}}+\|\bm\delta_i\|).
\end{align}
Note that in~\eqref{eq:diff_x}, the first inequality uses~\eqref{eq:up_x_2} and the non-expansiveness of projection, and the second inequality uses Fact~\ref{fact:cvx_comb}, Assumption~\ref{Asm:function} and~\eqref{eq:up_x_1}.

Also, observe that
\begin{align}
    f_i(\bx_i,\bz_n)-f_i(\bx_i^+,\bz_n)
    =&f_i(\bx_i,\bar{\bz})-f_i(\bx_i^+,\bar{\bz})+\frac{\gamma_i}{2}\left(\|\bx_i-\bz_n\|^2-\|\bx_i^+-\bz_n\|^2-\|\bx_i-\bar{\bz}\|^2+\|\bx_i^+-\bar{\bz}\|^2\right) \nonumber\\
    =&f_i(\bx_i,\bar{\bz})-f_i(\bx_i^+,\bar{\bz})+\gamma_i\left\langle\bx_i-\bx_i^+,\bar{\bz}-\bz_n\right\rangle \nonumber\\
    \leq&f_i(\bx_i,\bar{\bz})-f_i(\bx_i^+,\bar{\bz})+\gamma_i\left\|\bx_{i}-\bx_{i}^+\right\|\|\bar{\bz}-\bz_n\| \nonumber \\
\label{eq:sufficient_x_4}
    \leq&f_i(\bx_i,\bar{\bz})-f_i(\bx_i^+,\bar{\bz})+\gamma_i(\zeta D_{\mathcal{X}}+\eta_x\left(D_{g_i}+\gamma_i D_{\mathcal{X}}+\|\bm\delta_i\|\right))\|\bar{\bz}-\bz_n\|
\end{align}
where the first equality is due to the definition of $f_i$,
the second inequality uses~\eqref{eq:diff_x}.
Given~\eqref{eq:sufficient_x_1} and~\eqref{eq:sufficient_x_4}, we can see that the desired result holds.
The proof is complete.
$\hfill\blacksquare$

\paragraph*{$\bz_n$-block updates.}
Similarly, we turn to characterize the progress of one step of the $\bz_n$-block DGD update;
i.e., lines~\ref{alg:ca_1}-\ref{alg:ca_2} in Algorithm~\ref{alg:FedBCD_full}.
We will use the following temporary notations in this subsection for simplicity:
\[
    \bx_i=\bx_i^{(t+1)},~\bar{\bz}^+=\bar{\bz}^{(t+1)},~\bar{\bz}=\bar{\bz}^{(t)},~\bz_n=\bz_n^{(t)},~\bw_n=\bw_n^{(t)},~a_{n,m}=a_{n,m}^{(t)}.
\]
Given these notations, we can rewrite lines~\ref{alg:ca_1}-\ref{alg:ca_2} as
\begin{subequations}
\label{eq:suff_z1}
\begin{align}
    \bw_n &= \sum_{m\in\mathcal{V}}a_{n,m}\bz_m,\\
    \bz_n^+ &= \bw_n-\eta_z\sum_{i\in\mathcal{Q}_n}\nabla_{\bz_n}f_i(\bx_i,\bw_n) = \bw_n-\eta_z\sum_{i\in\mathcal{Q}_n}\gamma_i(\bw_n-\bx_i).
\end{align}
\end{subequations}
We have the following result.
\begin{Lemma}
\label{lemma:sufficient_decrease_z}
Suppose that Assumptions~\ref{Asm:function}-\ref{asm:stepsize} hold.
Then, the update $\bar{\bz}^+$ generated following~\eqref{eq:suff_z1} satisfies that
\begin{align*}
    &\frac{\eta_z}{2|\mathcal{V}|}\left\|\sum_{n\in\mathcal{V}}\sum_{i\in\mathcal{Q}_n}\nabla_{\bz_n}f_i(\bx_i,\bar{\bz})\right\|^2
    \leq \sum_{n\in\mathcal{V}}\sum_{i\in\mathcal{Q}_n}f_i(\bx_i,\bar{\bz})-\sum_{n\in\mathcal{V}}\sum_{i\in\mathcal{Q}_n}f_i(\bx_i,\bar{\bz}^+)+\frac{\eta_z\gamma^2|\mathcal{Q}|^2}{2}\sum_{n\in\mathcal{V}}\left\|\bar{\bz}-\bz_n\right\|^2.
\end{align*}
\end{Lemma}
{\it Proof:}
To start with, note that we have
\begin{align}
    \bar{\bz}^+-\bar{\bz}
    &= \frac{1}{|\mathcal{V}|}\left(\sum_{n\in\mathcal{V}}\bw_n-\eta_z\sum_{n\in\mathcal{V}}\sum_{i\in\mathcal{Q}_n}\nabla_{\bz_n}f_i(\bx_i,\bw_n)\right)-\bar{\bz} \nonumber\\
    &= \frac{1}{|\mathcal{V}|}\left(\sum_{m\in\mathcal{V}}\bz_m\sum_{n\in\mathcal{V}}a_{n,m}-\eta_z\sum_{n\in\mathcal{V}}\sum_{i\in\mathcal{Q}_n}\nabla_{\bz_n}f_i(\bx_i,\bw_n)\right)-\bar{\bz} \nonumber\\
    &= -\frac{\eta_z}{|\mathcal{V}|}\sum_{n\in\mathcal{V}}\sum_{i\in\mathcal{Q}_n}\nabla_{\bz_n}f_i(\bx_i,\bw_n)\nonumber,
\end{align}
By the Lipschitz smoothness of $f_i$ w.r.t. $\bz_n$ (see Fact~\ref{fact:lips}) and the above inequality, we have
\begin{align}
    &\sum_{n\in\mathcal{V}}\sum_{i\in\mathcal{Q}_n}f_i(\bx_i,\bar{\bz}^+)\nonumber\\
    \leq& \sum_{n\in\mathcal{V}}\sum_{i\in\mathcal{Q}_n}f_i(\bx_i,\bar{\bz})+\left\langle\sum_{n\in\mathcal{V}}\sum_{i\in\mathcal{Q}_n}\nabla_{\bz_n}f_i(\bx_i,\bar{\bz}),\bar{\bz}^+-\bar{\bz}\right\rangle+\sum_{n\in\mathcal{V}}\sum_{i\in\mathcal{Q}_n}\frac{\gamma_i}{2}\|\bar{\bz}^+-\bar{\bz}\|^2 \nonumber\\
    \leq& \sum_{n\in\mathcal{V}}\sum_{i\in\mathcal{Q}_n}f_i(\bx_i,\bar{\bz})-\frac{\eta_z}{|\mathcal{V}|}\left\langle\sum_{n\in\mathcal{V}}\sum_{i\in\mathcal{Q}_n}\nabla_{\bz_n}f_i(\bx_i,\bar{\bz}),\sum_{n\in\mathcal{V}}\sum_{i\in\mathcal{Q}_n}\nabla_{\bz_n}f_i(\bx_i,\bw_n)\right\rangle+\frac{\eta_z^2\gamma|\mathcal{Q}|}{2|\mathcal{V}|}\left\|\sum_{n\in\mathcal{V}}\sum_{i\in\mathcal{Q}_n}\nabla_{\bz_n}f_i(\bx_i,\bw_n)\right\|^2\nonumber\\
    =& \sum_{n\in\mathcal{V}}\sum_{i\in\mathcal{Q}_n}f_i(\bx_i,\bar{\bz})-\frac{\eta_z}{2|\mathcal{V}|}\left\|\sum_{n\in\mathcal{V}}\sum_{i\in\mathcal{Q}_n}\nabla_{\bz_n}f_i(\bx_i,\bar{\bz})\right\|^2-\frac{\eta_z}{2|\mathcal{V}|}\left\|\sum_{n\in\mathcal{V}}\sum_{i\in\mathcal{Q}_n}\nabla_{\bz_n}f_i(\bx_i,\bw_n)\right\|^2\nonumber\\
    &+\frac{\eta_z}{2|\mathcal{V}|}\left\|\sum_{n\in\mathcal{V}}\sum_{i\in\mathcal{Q}_n}\left(\nabla_{\bz_n}f_i(\bx_i,\bar{\bz})-\nabla_{\bz_n}f_i(\bx_i,\bw_n)\right)\right\|^2+\frac{\eta_z^2\gamma|\mathcal{Q}|}{2|\mathcal{V}|}\left\|\sum_{n\in\mathcal{V}}\sum_{i\in\mathcal{Q}_n}\nabla_{\bz_n}f_i(\bx_i,\bw_n)\right\|^2\nonumber\\
    =& \sum_{n\in\mathcal{V}}\sum_{i\in\mathcal{Q}_n}f_i(\bx_i,\bar{\bz})-\frac{\eta_z}{2|\mathcal{V}|}\left\|\sum_{n\in\mathcal{V}}\sum_{i\in\mathcal{Q}_n}\nabla_{\bz_n}f_i(\bx_i,\bar{\bz})\right\|^2+\frac{\eta_z}{2|\mathcal{V}|}\left\|\sum_{n\in\mathcal{V}}\sum_{i\in\mathcal{Q}_n}\gamma_i(\bar{\bz}-\bw_n)\right\|^2\nonumber\\
    &-\frac{\eta_z}{2|\mathcal{V}|}(1-\eta_z\gamma|\mathcal{Q}|)\left\|\sum_{n\in\mathcal{V}}\sum_{i\in\mathcal{Q}_n}\nabla_{\bz_n}f_i(\bx_i,\bw_n)\right\|^2\nonumber\\
\label{eq:sufficient_z_1}
    \leq& \sum_{n\in\mathcal{V}}\sum_{i\in\mathcal{Q}_n}f_i(\bx_i,\bar{\bz})-\frac{\eta_z}{2|\mathcal{V}|}\left\|\sum_{n\in\mathcal{V}}\sum_{i\in\mathcal{Q}_n}\nabla_{\bz_n}f_i(\bx_i,\bar{\bz})\right\|^2+\frac{\eta_z\gamma^2|\mathcal{Q}|^2}{2}\sum_{n\in\mathcal{V}}\left\|\bar{\bz}-\bw_n\right\|^2,
\end{align}
where the last equality uses $\nabla_{\bz_n}f_i(\bx_i,\bar{\bz})-\nabla_{\bz_n}f_i(\bx_i,\bw_n)=\gamma_i(\bar{\bz}-\bw_n)$, and the last inequality uses $1-\eta_z\gamma|\mathcal{Q}|\geq0$ (by Assumption~\ref{asm:stepsize}) and $\|\sum_{n\in\mathcal{V}}\sum_{i\in\mathcal{Q}_n}\bv_i\|^2\leq|\mathcal{V}||\mathcal{Q}|\sum_{n\in\mathcal{V}}\sum_{i\in\mathcal{Q}_n}\|\bv_i\|^2$.
Also notice that
\begin{align}
\label{eq:sufficient_z_2}
    \sum_{n\in\mathcal{V}}\|\bw_n-\bar{\bz}\|^2=\sum_{n\in\mathcal{V}}\left\|\sum_{m\in\mathcal{V}}a_{n,m}\bz_m-\bar{\bz}\right\|^2\leq\sum_{n\in\mathcal{V}}\sum_{m\in\mathcal{V}}a_{n,m}\|\bz_m-\bar{\bz}\|^2=\sum_{m\in\mathcal{V}}\|\bz_m-\bar{\bz}\|^2,
\end{align}
where the inequality uses $\|\sum_{n\in\mathcal{V}}a_n\bv_n\|^2\leq\sum_{n\in\mathcal{V}}a_n\|\bv_n\|^2$ for any $a_n\geq0$ such that $\sum_{n\in\mathcal{V}}a_n=1$.
Combing~\eqref{eq:sufficient_z_1} and~\eqref{eq:sufficient_z_2}, we obtain the desired result.
The proof is complete.
$\hfill\blacksquare$

\subsection{Proof of Theorem~\ref{thm}}
To start with, we need to introduce two lemmas that provide us with key upper bounds that pave the way for our final analysis.
\begin{Lemma}
\label{lemma:z_consensus}
Suppose that Assumptions~\ref{Asm:subset}-\ref{asm:stepsize} hold.
Define the below constants
\begin{align*}
    C_1 &= \theta|\mathcal{V}|\frac{1-\beta^T}{1-\beta}\cdot\max_{m\in\mathcal{V}}\|\bz_m^{(0)}\|,\quad
    C_2 = 2\gamma D_{\mathcal{X}}|\mathcal{Q}|+\frac{\gamma\theta D_{\mathcal{X}}|\mathcal{Q}||\mathcal{V}|}{1-\beta},\\
    C_3 &= \theta^2|\mathcal{V}|^2\frac{1-\beta^{2T}}{1-\beta^2}\cdot\max_{m\in\mathcal{V}}\|\bz_m^{(0)}\|^2,\quad
    C_4 = 2\theta|\mathcal{V}|\gamma D_{\mathcal{X}}|\mathcal{Q}|\frac{\theta|\mathcal{V}|}{1-\beta^2}\cdot\max_{m\in\mathcal{V}}\|\bz_m^{(0)}\|,\\
    C_5 &= 2\theta|\mathcal{V}|\gamma D_{\mathcal{X}}|\mathcal{Q}|\frac{1-\beta^T}{1-\beta}\cdot \max_{m\in\mathcal{V}}\|\bz_m^{(0)}\|,\quad
    C_6 = \gamma^2\mathcal{D}_{\mathcal{X}}|\mathcal{Q}|\left(\frac{\theta^2|\mathcal{V}|^2}{1-\beta^2}+\left(\frac{2\beta\theta^2|\mathcal{V}|^2}{1-\beta}+4\theta|\mathcal{V}|\right)\frac{1}{1-\beta}+4\right),
\end{align*}
where
\[
    \theta=\left(1-\frac{c}{4|\mathcal{V}|^2}\right)^{-2},\quad \beta=\left(1-\frac{c}{4|\mathcal{V}|^2}\right)^{1/q}.
\]
Note that $c$ is from Assumption~\ref{Asm:5} and $q$ is from Assumption~\ref{Asm:subset}.
Then, for all $t$ and $n$, the sequences $\{\bar{\bz}^{(t)}\}$ and $\{\bz_n^{(t)}\}$ generated by Algorithm~2 has the below upper bounds
\begin{align*}
    \sum_{t=0}^{T-1}\|\bar{\bz}^{(t)}-\bz_n^{(t)}\| &\leq {\rm C}_1+\eta_z T C_2\\
    \sum_{t=0}^{T-1}\|\bar{\bz}^{(t)}-\bz_n^{(t)}\|^2 &\leq C_3 + \eta_z TC_4 + \eta_z C_5 + \eta_z^2 T C_6.
\end{align*}
\end{Lemma}
\begin{Lemma}
\label{lemma:accumulation}
Suppose that Assumptions~\ref{Asm:subset}-\ref{asm:batch_size} hold and the batch size number is $R$.
Define the below constants
\begin{align*}
    &C_7 = 2\gamma|\mathcal{V}||\mathcal{Q}|(\zeta D_{\mathcal{X}}/\eta_x+D_{g_i}+\gamma D_{\mathcal{X}}+\sigma/\sqrt{R})C_1+\gamma^2|\mathcal{V}||\mathcal{Q}|^2C_3 + \eta_z\gamma^2|\mathcal{V}||\mathcal{Q}|^2C_5,\\
    &C_8 = 2\gamma|\mathcal{V}||\mathcal{Q}|(\zeta D_{\mathcal{X}}/\eta_x+D_{g_i}+\gamma D_{\mathcal{X}}+\sigma/\sqrt{R})C_2+\gamma^2|\mathcal{V}||\mathcal{Q}|^2C_4,\\
    &C_9 = \gamma^2|\mathcal{V}||\mathcal{Q}|^2C_6,\\
    &C_{10} = 2|\mathcal{V}||\mathcal{Q}|(\zeta D_{\mathcal{X}}/\eta_x+D_{g_i}+\gamma D_{\mathcal{X}}),
\end{align*}
where $C_1$-$C_6$ are given in Lemma~\ref{lemma:z_consensus}.
Let the communication round number $T$ be sufficiently large such that
\begin{equation}
\label{eq:T_upper}
    \sqrt{T} \geq \max\left\{\frac{\Delta^*+\eta_{\sf max}(C_7+d^2C_9)}{\eta_{\sf max}dC_8},\frac{C_1}{dC_2}\right\},
\end{equation}
where $\eta_{\sf max} = \max\{\eta_z,\eta_x\}$.
Then, for all $i,n$, the sequences $\{\bx_i^{(t)}\}$ and $\{\bar{\bz}^{(t)}\}$ generated by Algorithm~\ref{alg:FedBCD_full} satisfy that
\begin{align*}
    &\frac{1}{T}\sum_{t=0}^{T-1}\mathbb{E}\left[\left\|\sum_{n\in\mathcal{V}}\sum_{i\in\mathcal{Q}_n}\nabla_{\bz_n}f_i(\bx_i^{(t+1)},\bar{\bz}^{(t)})\right\|^2\right] \leq\frac{B_1}{\sqrt{T}}+\frac{B_2}{\sqrt{R}},\\
    &\frac{1}{T}\sum_{t=0}^{T-1}\mathbb{E}\left[\frac{1}{\eta_x^2}\left\|\bx_i^{(t+1)}-\bx_i^{(t)}\right\|^2\right] \leq \frac{B_1}{\sqrt{T}}+\frac{B_2}{\sqrt{R}}.
\end{align*}
where $B_1 = \frac{2\eta_{\sf max}dC_8}{\min\{1/|\mathcal{V}|,1-\omega^2\}\eta_{\sf min}}, B_2 = \frac{\eta_{\sf max}(2\sigma^2/\sqrt{R}+\sigma C_{10})}{\min\{1/|\mathcal{V}|,1-\omega^2\}\eta_{\sf min}}$, $\eta_{\sf min} = \min\{\eta_z,\eta_x\}$.
\end{Lemma}
The proof of Lemmas~\ref{lemma:z_consensus}-\ref{lemma:accumulation} are given in Sections~\ref{proof:lemma5}-\ref{proof:lemma6}.

From Lemmas~\ref{lemma:z_consensus}-\ref{lemma:accumulation}, we already have
\begin{align}
\label{eq:thm_f1}
    \frac{1}{T}\sum_{t=0}^{T-1}\mathbb{E}\left[\left\|\sum_{n\in\mathcal{V}}\sum_{i\in\mathcal{Q}_n}\nabla_{\bz_n}f_i(\bx_i^{(t+1)},\bar{\bz}^{(t)})\right\|^2\right] &\leq \frac{B_1}{\sqrt{T}}+\frac{B_2}{\sqrt{R}},\\
\label{eq:thm_f2}
    \frac{1}{T}\sum_{t=0}^{T-1}\|\bar{\bz}^{(t)}-\bz_n^{(t)}\|&\leq\frac{C}{\sqrt{T}},
\end{align}
where $C=2dC_2$ (recall the requirement on $T$ in~\eqref{eq:T_upper}).
We still need to show the first inequality in Theorem~\ref{thm}(i).
Let $t_i$ be the element in the subsequence $\{T_i\}\subseteq\{0,1,\ldots,T-1\}$ such that $i\in\mathcal{Q}_n^{(t_i)}$ for all $t_i$.
Again, we can simplify the notation as
\begin{align}
\label{eq:notationss}
    &\quad\bx_i^+=\bx_i^{(t_i+1)},~\bx_i=\bx_i^{(t_i)},~\bx_{i,\sf ex}=\bx_{i,\sf ex}^{(t_i)},~\bz_n=\bz_n^{(t_i)},~\bar{\bz}=\bar{\bz}^{(t_i)},~\bm\delta_i=\bm\delta_i^{(t_i)}.
\end{align}
Recall that from Fact~\ref{fact:1} we have
\begin{equation}
\label{eq:thm_2}
    \left\langle\nabla_{\bx_i}f_i(\bx_{i,\sf ex},\bz_n)+\bm\delta_{i}+\frac{1}{\eta_x}\left(\bx_i^+-\bx_{i,\sf ex}\right),\hat{\bx}_i-\bx_i^+\right\rangle\geq 0,~\forall\hat{\bx}_i\in\mathcal{X}.
\end{equation}
Also notice that
\begin{align}
    &\left\langle\nabla_{\bx_i}f_i(\bx_{i,\sf ex},\bz_n)+\bm\delta_i+\frac{1}{\eta_x}\left(\bx_i^+-\bx_{i,\sf ex}\right)-\nabla_{\bx_i}f_i(\bx_i^+,\bar{\bz}),\hat{\bx}_i-\bx_i^+\right\rangle \nonumber\\
    \leq&\left(\left\|\nabla_{\bx_i}f_i(\bx_{i,\sf ex},\bz_n)-\nabla_{\bx_i}f_i(\bx_{i}^+,\bar{\bz})\right\|+\frac{1}{\eta_x}\left\|\bx_i^+-\bx_{i,\sf ex}\right\|\right)D_{\mathcal{X}}+\|\bm\delta_i\|D_{\mathcal{X}} \nonumber\\
    =&\left(\left\|\nabla g_i(\bx_{i,\sf ex})-\nabla g_i(\bx_i^+)+\gamma_i(\bx_{i,\sf ex}-\bx_i^+)-\gamma_i(\bz_n-\bar{\bz})\right\|+\frac{1}{\eta_x}\left\|\bx_i^+-\bx_{i,\sf ex}\right\|\right)D_{\mathcal{X}}+\|\bm\delta_i\|D_{\mathcal{X}} \nonumber\\
    \leq&\left(\left(L_i+\gamma_i+\frac{1}{\eta_x}\right)\left\|\bx_i^+-\bx_{i,\sf ex}\right\|+\gamma_i\|\bar{\bz}-\bz_n\|\right)D_{\mathcal{X}}+\|\bm\delta_i\|D_{\mathcal{X}} \nonumber\\
\label{eq:thm_3}
    \leq&\left(\left(L_i+\gamma_i+\frac{1}{\eta_x}\right)(\|\bx_i^+-\bx_i\|+\zeta ||\bx_i-\bx_i^-||)+\gamma_i\|\bar{\bz}-\bz_n\|\right)D_{\mathcal{X}}+\|\bm\delta_i\|D_{\mathcal{X}}.
\end{align}
Combining~\eqref{eq:thm_2} and~\eqref{eq:thm_3}, we have
\begin{align}
\label{eq:final}
    \left\langle\nabla_{\bx_i}f_i(\bx_i^+,\bar{\bz}),\hat{\bx}_i-\bx_i^+\right\rangle
    \geq&-\left(\left(L_i+\gamma_i+\frac{1}{\eta_x}\right)(\|\bx_i^+-\bx_i\|+\zeta ||\bx_i-\bx_i^-||)+\gamma_i\|\bar{\bz}-\bz_n\|\right)D_{\mathcal{X}}-\|\bm\delta_i\|D_{\mathcal{X}}.
\end{align}
Substituting~\eqref{eq:notationss} to~\eqref{eq:final}, we obtain
\begin{align}
    &\frac{1}{T}\sum_{t=0}^{T-1}\mathbb{E}\left[\left\langle\nabla_{\bx_i}f_i(\bx_i^{(t+1)},\bar{\bz}^{(t)}),\hat{\bx}_i-\bx_i^{(t+1)}\right\rangle\right]\nonumber\\
    \geq&-2D_{\mathcal{X}}\frac{1}{T}\sum_{t=0}^{T-1}\mathbb{E}\left[\frac{1}{\eta_x}\left\|\bx_i^{(t+1)}-\bx_{i}^{(t)}\right\|\right]-\gamma_iD_{\mathcal{X}}\frac{1}{T}\sum_{t=0}^{T-1}\mathbb{E}\left[\|\bar{\bz}^{(t)}-\bz_n^{(t_i)}\|\right]-\frac{\sigma D_{\mathcal{X}}}{\sqrt{R}}-2\zeta D_{\mathcal{X}}\frac{1}{T}\sum_{t=0}^{T-1}\mathbb{E}\left[\frac{1}{\eta_x}\left\|\bx_i^{(t)}-\bx_{i}^{(t-1)}\right\|\right]\nonumber\\
    \geq&-4D_{\mathcal{X}}\frac{1}{T}\sum_{t=0}^{T}\mathbb{E}\left[\frac{1}{\eta_x}\left\|\bx_i^{(t)}-\bx_{i}^{(t-1)}\right\|\right]-\frac{\gamma D_{\mathcal{X}}C}{\sqrt{T}}-\frac{\sigma D_{\mathcal{X}}}{\sqrt{R}}
    \nonumber\\
    \geq& -\frac{4D_{\mathcal{X}}(T+1)}{T}\sqrt{\frac{B_1}{\sqrt{T+1}}+\frac{B_2}{\sqrt{R}}}-\frac{\gamma D_{\mathcal{X}} C}{\sqrt{T}}-\frac{\sigma D_{\mathcal{X}}}{\sqrt{R}}\nonumber\\
    \geq& -\frac{4D_{\mathcal{X}}(T+1)}{T}\left(\sqrt{\frac{B_1}{\sqrt{T+1}}}+\sqrt{\frac{B_2}{\sqrt{R}}}\right)-\frac{\gamma D_{\mathcal{X}} C}{\sqrt{T}}-\frac{\sigma D_{\mathcal{X}}}{\sqrt{R}}\nonumber\\
\label{eq:thm_f}
    \geq& -\frac{A_1}{T^{1/4}}-\frac{A_2}{R^{1/4}},
\end{align}
where the third inequality is based on Lemma~\ref{lemma:accumulation}, the forth inequality applies that $\sqrt{x+y}\leq \sqrt{x}+\sqrt{y}$ for positive $x,y$, and the last inequality can be achieved by using suitable $A_1$ and $A_2$.
Given~\eqref{eq:thm_f1},~\eqref{eq:thm_f2} and~\eqref{eq:thm_f}, we obtain the result in Theorem~\ref{thm}(i).

If the exact gradients are used in FedBCD, we will have that $\delta_i^{(t)}=\bm0$ for all $i,t$.
By removing all terms related to $\delta_i^{(t)}=\bm0$ in the above derivation, we will finally show Theorem~\ref{thm}(ii);
i.e.,
\begin{align*}
    &\frac{1}{T}\sum_{t=0}^{T-1}\mathbb{E}\left[\left\langle\nabla_{\bx_i}f_i(\bx_i^{(t+1)},\bar{\bz}^{(t)}),\hat{\bx}_i-\bx_i^{(t+1)}\right\rangle\right] \geq -\frac{A_1}{T^{1/4}},~\forall \hat{\bx}\in\mathcal{X},~i\in\mathcal{Q}_n,n\in\mathcal{V},\\
    &\frac{1}{T}\sum_{t=0}^{T-1}\mathbb{E}\left[\left\|\sum_{n\in\mathcal{V}}\sum_{i\in\mathcal{Q}_n}\nabla_{\bz_n}f_i(\bx_i^{(t+1)},\bar{\bz}^{(t)})\right\|^2\right] \leq \frac{B_1}{\sqrt{T}},\\        &\frac{1}{T}\sum_{t=0}^{T-1}\|\bar{\bz}^{(t)}-\bz_n^{(t)}\|\leq\frac{C}{\sqrt{T}},~\forall n\in\mathcal{V}.
\end{align*}
We complete the overall proof.

\subsection{Proof of Lemma~\ref{lemma:z_consensus}}
\label{proof:lemma5}
For any integer $t$ and $n$, define
\begin{equation}
\label{eq:Lemma2_1}
    \bp_{n}^{(t+1)} = \bz_n^{(t+1)}-\sum_{m\in\mathcal{V}}a_{n,m}^{(t)}\bz_m^{(t)}.
\end{equation}
Rearranging~\eqref{eq:Lemma2_1}, we observe that
\begin{align*}
    \bz_n^{(t+1)}
    =& \bp_n^{(t+1)}+\sum_{m\in\mathcal{V}}a_{n,m}^{(t)}\bz_m^{(t)}\\
    =& \bp_n^{(t+1)}+\sum_{m\in\mathcal{V}}a_{n,m}^{(t)}\left(\bp_m^{(t)}+\sum_{\ell\in\mathcal{V}}a_{m,\ell}^{(t-1)}\bz_\ell^{(t-1)}\right)\\
    =&\bp_n^{(t+1)} + \sum_{m\in\mathcal{V}}[\Phi(t,t-1)]_{n,m}\bp_m^{(t)} + \sum_{\ell\in\mathcal{V}}\bz_\ell^{(t-1)}\sum_{m\in\mathcal{V}}a_{n,m}^{(t)}a_{m,\ell}^{(t-1)}\\
    =&\bp_n^{(t+1)} + \sum_{m\in\mathcal{V}}[\Phi(t,t-1)]_{n,m}\bp_m^{(t)}+\sum_{m\in\mathcal{V}}[\Phi(t,t-2)]_{n,m}\bz_m^{(t-1)}.
\end{align*}
Repeating the above rearrangement, we will finally obtain
\begin{equation}
\label{eq:Lemma2_2}
    \bz_n^{(t+1)}=\sum_{m\in\mathcal{V}}[\Phi(t,-1)]_{n,m}\bz_m^{(0)}+\bp_n^{(t+1)}+\sum_{\ell=1}^{t}\left(\sum_{m\in\mathcal{V}}[\Phi(t,\ell-1)]_{n,m}\bp_m^{(\ell)}\right).
\end{equation}
Based on~\eqref{eq:Lemma2_1}, we can also rewrite $\bar{\bz}^{(t+1)}$ as
\begin{align*}
    \bar{\bz}^{(t+1)}
    &= \frac{1}{|\mathcal{V}|}\left(\sum_{n\in\mathcal{V}}\bp_n^{(t+1)}+\sum_{n\in\mathcal{V}}\sum_{m\in\mathcal{V}}a_{n,m}^{(t)}\bz_m^{(t)}\right)
    = \frac{1}{|\mathcal{V}|}\left(\sum_{n\in\mathcal{V}}\bp_n^{(t+1)}+\sum_{m\in\mathcal{V}}\bz_m^{(t)}\right)= \bar{\bz}^{(t)}+\frac{1}{|\mathcal{V}|}\sum_{n\in\mathcal{V}}\bp_n^{(t+1)}.
\end{align*}
It follows that
\begin{equation}
\label{eq:Lemma2_3}
    \bar{\bz}^{(t+1)} = \bar{\bz}^{(0)}+\frac{1}{|\mathcal{V}|}\sum_{\ell=1}^{t+1}\sum_{n\in\mathcal{V}}\bp_n^{(\ell)} = \frac{1}{|\mathcal{V}|}\sum_{n\in\mathcal{V}}\bz_n^{(0)}+\frac{1}{|\mathcal{V}|}\sum_{\ell=1}^{t+1}\sum_{n\in\mathcal{V}}\bp_n^{(\ell)}.
\end{equation}
Given~\eqref{eq:Lemma2_2} and~\eqref{eq:Lemma2_3}, we have
\begin{align*}
    &\|\bar{\bz}^{(t+1)}-\bz_n^{(t+1)}\|\\
    =& \left\|\frac{1}{|\mathcal{V}|}\sum_{m\in\mathcal{V}}\bz_m^{(0)}+\frac{1}{|\mathcal{V}|}\sum_{\ell=1}^{t+1}\sum_{m\in\mathcal{V}}\bp_m^{(\ell)}-\sum_{m\in\mathcal{V}}[\Phi(t,-1)]_{n,m}\bz_m^{(0)}-\bp_n^{(t+1)}-\sum_{\ell=1}^t\left(\sum_{m\in\mathcal{V}}[\Phi(t,\ell-1)]_{n,m}\bp_m^{(\ell)}\right)\right\|\\
    \leq&\sum_{m\in\mathcal{V}}\left|\frac{1}{|\mathcal{V}|}-[\Phi(t,-1)]_{n,m}\right|\left\|\bz_m^{(0)}\right\|+\sum_{\ell=1}^t\sum_{m\in\mathcal{V}}\left|\frac{1}{|\mathcal{V}|}-[\Phi(t,\ell-1)]_{n,m}\right|\left\|\bp_m^{(\ell)}\right\|+\frac{1}{|\mathcal{V}|}\sum_{m\in\mathcal{V}}\left\|\bp_m^{(t+1)}\right\|+\left\|\bp_n^{(t+1)}\right\|.
\end{align*}
To proceed, we need to resort to the following known result.
\begin{Lemma}[Matrix Convergence~{\cite{ram2010distributed}}]
\label{Lemma:matrix_convergence}
    Suppose that Assumptions~\ref{Asm:subset} and \ref{Asm:5} hold.
    Then we have that
        \[
            \left|[\bm\Phi(t,s)]_{n,m}-\frac{1}{|\mathcal{V}|}\right|\leq\theta\beta^{t-s},
        \]
        where
        \[
            \theta=\left(1-\frac{c}{4|\mathcal{V}|^2}\right)^{-2},\quad \beta=\left(1-\frac{c}{4|\mathcal{V}|^2}\right)^{1/q}.
        \]
    Note that $c$ is from Assumption~\ref{Asm:5} and $q$ is from Assumption~\ref{Asm:subset}.
\end{Lemma}
Using Lemma~\ref{Lemma:matrix_convergence}, we have
\begin{align}
\label{eq:Lemma2_4}
    \left\|\bar{\bz}^{(t+1)}-\bz_n^{(t+1)}\right\|
    \leq& \theta\beta^{t+1}|\mathcal{V}|\max_{m\in\mathcal{V}}\left\|\bz_m^{(0)}\right\|+\sum_{\ell=1}^t\theta\beta^{t+1-\ell}\sum_{m\in\mathcal{V}}\left\|\bp_m^{(\ell)}\right\|+\frac{1}{|\mathcal{V}|}\sum_{m\in\mathcal{V}}\left\|\bp_m^{(t+1)}\right\|+\left\|\bp_n^{(t+1)}\right\|.
\end{align}
The next task is to bound $\|\bp_n^{(t)}\|$.
We have
\begin{align}
\label{eq:Lemma2_5}
    \left\|\bp_n^{(t+1)}\right\|
    =\left\|\bz_n^{(t+1)}-\sum_{m\in\mathcal{V}}a_{n,m}^{(t)}\bz_m^{(t)}\right\|
    \leq \eta_z^{(t+1)}\sum_{i\in\mathcal{Q}_n}\gamma_i\left\|\bw_n^{(t)}-\bx_i^{(t+1)}\right\|
    \leq \eta_z^{(t+1)}{\gamma}D_{\mathcal{X}}|{\mathcal{Q}}|,
\end{align}
where the second inequality uses $\bw_n^{(t)}\in\mathcal{X}$ and $\bx_i^{(t+1)}\in\mathcal{X}$ (see Fact~\ref{fact:cvx_comb}).
Combining~\eqref{eq:Lemma2_4} and~\eqref{eq:Lemma2_5}, and rearranging the corresponding terms, we obtain
\begin{equation*}
    \|\bar{\bz}^{(t+1)}-\bz_n^{(t+1)}\|\leq \theta\beta^{t+1}|\mathcal{V}|\max_{m\in\mathcal{V}}\|\bz_m^{(0)}\|+{\gamma} D_{\mathcal{X}}|{\mathcal{Q}}|\left(\theta|\mathcal{V}|\sum_{\ell=1}^t\beta^{t+1-\ell}\eta_z+2\eta_z\right).
\end{equation*}
{\bf Hint}: To extend the above result to the case where $K_z>1$, notice that above results always hold no matter how $\bx_i$'s are updated (as long as they lie on $\mathcal{X}$).
In view of this, we can, loosely speaking, ``unfold'' the inner loop w.r.t. $k$ and the outer loop w.r.t. $t$ into a single loop.
The above analysis remains unchanged for the resulting new loop.

Using the above result, we obtain the first result
\begin{align}
    \sum_{t=0}^{T-1}\|\bar{\bz}^{(t)}-\bz_n^{(t)}\|
    &\leq \sum_{t=0}^{T-1}\left(\theta\beta^{t}|\mathcal{V}|\max_{m\in\mathcal{V}}\|\bz_m^{(0)}\|+{\gamma} D_{\mathcal{X}}|{\mathcal{Q}}|\left(\theta|\mathcal{V}|\sum_{\ell=1}^{t-1}\beta^{t-\ell}\eta_z+2\eta_z\right)\right) \notag\\
    &\leq \theta|\mathcal{V}|\max_{m\in\mathcal{V}}\|\bz_m^{(0)}\|\frac{1-\beta^T}{1-\beta}+ \eta_z\left(2T\gamma D_{\mathcal{X}}|\mathcal{Q}|+\gamma\theta D_{\mathcal{X}}|\mathcal{Q}||\mathcal{V}|\sum_{t=0}^{T-1}\sum_{\ell=0}^{t}\beta^{t-\ell}\right) \notag \\
    &\leq \theta|\mathcal{V}|\max_{m\in\mathcal{V}}\|\bz_m^{(0)}\|\frac{1-\beta^T}{1-\beta}+ \eta_z T\left(2\gamma D_{\mathcal{X}}|\mathcal{Q}|+\frac{\gamma\theta D_{\mathcal{X}}|\mathcal{Q}||\mathcal{V}|}{1-\beta}\right)\notag\\
    &= C_1+\eta_z T C_2, \notag
\end{align}
where $C_1$ and $C_2$ are defined in Lemma~\ref{lemma:z_consensus}, and the last inequality uses the result in \cite[Eq.~(18)]{pmlr-v97-assran19a}
\[
  \sum_{t=0}^{T-1}\sum_{\ell=0}^t\beta^{t-\ell}\leq \frac{T}{1-\beta},~~\forall\beta\in(0,1).
\]

To show the second result, we first have
\begin{align}
    \|\bar{\bz}^{(t)}-\bz_n^{(t)}\|^2
    \leq &\left(\underbrace{\theta^2\beta^{2t}|\mathcal{V}|^2\max_{m\in\mathcal{V}}\|\bz_m^{(0)}\|^2}_{a^{(t)}}+\underbrace{2\theta\beta^t|\mathcal{V}|\max_{m\in\mathcal{V}}\|\bz_m^{(0)}\|\gamma D_{\mathcal{X}}|\mathcal{Q}|\left(\theta|\mathcal{V}|\sum_{\ell=1}^{t-1}\beta^{t-\ell}\eta_z+2\eta_z\right)}_{b^{(t)}}+\right.\nonumber\\
\label{eq:lemma4_abc}
    &\left.\underbrace{\gamma^2 D_{\mathcal{X}}^2|\mathcal{Q}|^2\left(\theta|\mathcal{V}|\sum_{\ell=1}^{t-1}\beta^{t-\ell}\eta_z+2\eta_z\right)^2}_{c^{(t)}}\right).
\end{align}
It can be verified that
\begin{align}
\label{eq:lemma4_2_a}
    \sum_{t=0}^{T-1}a^{(t)} \leq \theta^2|\mathcal{V}|^2\max_{m\in\mathcal{V}}\|\bz_m^{(0)}\|^2\frac{1-\beta^{2T}}{1-\beta^2}.
\end{align}
\begin{align}
\label{eq:lemma4_2_b}
    \sum_{t=0}^{T-1}b^{(t)}
    &\leq
    2\eta_z\theta|\mathcal{V}|\max_{m\in\mathcal{V}}\|\bz_m^{(0)}\|\gamma D_{\mathcal{X}}|\mathcal{Q}|\left(\theta|\mathcal{V}|\sum_{t=0}^{T-1}\beta^t\sum_{\ell=0}^{t}\beta^{t-\ell}+\frac{1-\beta^T}{1-\beta}\right)\nonumber\\
    &\leq 2\eta_z\theta|\mathcal{V}|\max_{m\in\mathcal{V}}\|\bz_m^{(0)}\|\gamma D_{\mathcal{X}}|\mathcal{Q}|\left(\frac{T\theta|\mathcal{V}|}{1-\beta^2}+\frac{1-\beta^T}{1-\beta}\right),
\end{align}
where we use the result in \cite[Eq.~(19)]{pmlr-v97-assran19a}
\[
  \sum_{t=0}^{T-1}\beta^t\sum_{\ell=0}^t\beta^{t-\ell}\leq \frac{T}{1-\beta^2},~~\forall\beta\in(0,1).
\]
We can write $c^{(t)}$ as
\begin{align}
    c^{(t)}
    &= \eta_z^2\gamma^2 D_{\mathcal{X}}^2|\mathcal{Q}|^2\left(\theta^2|\mathcal{V}|^2\left(\sum_{\ell=1}^{t-1}\beta^{t-\ell}\right)^2+4\theta|\mathcal{V}|\sum_{\ell=1}^{t-1}\beta^{t-\ell}+4\right)\nonumber\\
    &= \eta_z^2\gamma^2 D_{\mathcal{X}}^2|\mathcal{Q}|^2\left(\theta^2|\mathcal{V}|^2\left(\sum_{\ell=1}^{t-1}\beta^{2(t-\ell)}+2\sum_{\ell=1}^{t-2}\sum_{j=\ell+1}^{t-1}\beta^{2t-\ell-j}\right)+4\theta|\mathcal{V}|\sum_{\ell=1}^{t-1}\beta^{t-\ell}+4\right) \nonumber\\
    &\leq \eta_z^2\gamma^2 D_{\mathcal{X}}^2|\mathcal{Q}|^2\left(\theta^2|\mathcal{V}|^2\left(\sum_{\ell=1}^{t-1}\beta^{2(t-\ell)}+2\sum_{\ell=1}^{t-2}\sum_{j=1}^{t-1}\beta^{2t-\ell-j}\right)+4\theta|\mathcal{V}|\sum_{\ell=1}^{t-1}\beta^{t-\ell}+4\right) \nonumber\\
    &= \eta_z^2\gamma^2 D_{\mathcal{X}}^2|\mathcal{Q}|^2\left(\theta^2|\mathcal{V}|^2\left(\sum_{\ell=1}^{t-1}\beta^{2(t-\ell)}+2\frac{\beta-\beta^t}{1-\beta}\sum_{\ell=1}^{t-2}\beta^{t-\ell}\right)+4\theta|\mathcal{V}|\sum_{\ell=1}^{t-1}\beta^{t-\ell}+4\right) \nonumber\\
    &\leq \eta_z^2\gamma^2 D_{\mathcal{X}}^2|\mathcal{Q}|^2\left(\theta^2|\mathcal{V}|^2\left(\sum_{\ell=1}^{t-1}\beta^{2(t-\ell)}+\frac{2\beta}{1-\beta}\sum_{\ell=1}^{t-1}\beta^{t-\ell}\right)+4\theta|\mathcal{V}|\sum_{\ell=1}^{t-1}\beta^{t-\ell}+4\right) \nonumber
\end{align}
It follows that
\begin{align}
    \sum_{t=0}^{T-1}c^{(t)}
    &\leq\eta_z^2\gamma^2\mathcal{D}_{\mathcal{X}}|\mathcal{Q}|\left(\theta^2|\mathcal{V}|^2\sum_{t=0}^{T-1}\sum_{\ell=0}^t\beta^{2(t-\ell)}+\left(\frac{2\beta\theta^2|\mathcal{V}|^2}{1-\beta}+4\theta|\mathcal{V}|\right)\sum_{t=0}^{T-1}\sum_{\ell=0}^t\beta^{t-\ell}+4T\right) \nonumber\\
\label{eq:lemma4_2_c}
    &\leq \eta_z^2T\gamma^2\mathcal{D}_{\mathcal{X}}|\mathcal{Q}|\left(\frac{\theta^2|\mathcal{V}|^2}{1-\beta^2}+\left(\frac{2\beta\theta^2|\mathcal{V}|^2}{1-\beta}+4\theta|\mathcal{V}|\right)\frac{1}{1-\beta}+4\right).
\end{align}
Combining~\eqref{eq:lemma4_abc}-\eqref{eq:lemma4_2_c}, we obtain
\begin{align*}
    \sum_{t=0}^{T-1}\|\bar{\bz}^{(t)}-\bz_n^{(t)}\|^2 \leq C_3 + \eta_z TC_4 + \eta_z C_5 + \eta_z^2 T C_6,
\end{align*}
where $C_3, C_4, C_5$ and $C_6$ are given in Lemma~\ref{lemma:z_consensus}.
We complete the proof.

\subsection{Proof of Lemma~\ref{lemma:accumulation}}
\label{proof:lemma6}
To start with, we will use the following simplified notations:
\begin{align*}
    \Delta_{x}^{(t)} &=
    \sum_{n\in\mathcal{V}}\sum_{i\in\mathcal{Q}_n}2\left(\mathbb{E}\left[f_i(\bx_i^{(t)},\bar{\bz}^{(t)})\right]-\mathbb{E}\left[f_i(\bx_i^{(t+1)},\bar{\bz}^{(t)})\right]\right)
    ,\\
    \Delta_{z}^{(t)} &= \sum_{n\in\mathcal{V}}\sum_{i\in\mathcal{Q}_n}2\left(\mathbb{E}\left[f_i(\bx_i^{(t+1)},\bar{\bz}^{(t)})\right]-\mathbb{E}\left[f_i(\bx_i^{(t+1)},\bar{\bz}^{(t+1)})\right]\right),\\
    e^{(t)}_{x_i} &= \begin{cases}\frac{1}{\eta_x}\mathbb{E}[\|\bx_i^{(t+1)}-\bx_i^{(t)}\|^2]-\zeta^2\left(\frac{1}{\eta_x}+\rho_i\right)\mathbb{E}[\|\bx_i^{(t)}-\bx_i^{(t^-)}\|^2], & i\in\mathcal{Q}_n^{(t)}\\0, & {\rm otherwise}\end{cases}
    ,\\
    e^{(t)}_z &= \frac{1}{|\mathcal{V}|}\mathbb{E}\left[\left\|\sum_{n\in\mathcal{V}}\sum_{i\in\mathcal{Q}_n}\nabla_{\bz_n}f_i(\bx_i^{(t+1)},\bar{\bz}^{(t)})\right\|^2\right],
\end{align*}
where $\bx_i^{(t^-)}$ denotes the last active update that is closest to $\bx_i^{(t)}$.
Invoking Lemma~\ref{Lemma:sufficient_decrease_xi} and Lemma~\ref{lemma:sufficient_decrease_z}, we have that
\begin{align}
\label{eq:main_t1}
    \sum_{n\in\mathcal{V}}\sum_{i\in\mathcal{Q}_n}\left(e_{x_i}^{(t)}-C_{x_i}^{(t)}\right)\leq \Delta_x^{(t)},\quad
    \eta_z\left(e_z^{(t)}-C_z^{(t)}\right)\leq \Delta_z^{(t)},
\end{align}
where
\begin{align*}
    C_{x_i}^{(t)} &= \begin{cases}2\mathbb{E}[(\zeta D_{\mathcal{X}}+\eta_x(D_{g_i}+\gamma_i D_{\mathcal{X}}+\|\bm\delta_i^{(t)}\|))(\gamma_i\|\bar{\bz}^{(t)}-\bz_n^{(t)}\|+\|\bm\delta_i^{(t)}\|)], & i\in\mathcal{Q}_n^{(t)}\\0, & {\rm otherwise}\end{cases},
    \\
    C_z^{(t)} &= \gamma^2|\mathcal{Q}|^2\sum_{n\in\mathcal{V}}\mathbb{E}\left[\|\bar{\bz}^{(t)}-\bz_n^{(t)}\|^2\right].
\end{align*}

Summing~\eqref{eq:main_t1} up, it can be checked that
\begin{align}
\label{eq:main_ttt}
    \sum_{t=0}^{T-1}\left(\sum_{n\in\mathcal{V}}\sum_{i\in\mathcal{Q}_n}(e_{x_i}^{(t)}-C_{x_i}^{(t)})+\eta_z\left(e_z^{(t)}-C_z^{(t)}\right)\right)
    \leq \Delta_T,
\end{align}
where $\Delta_T=\sum_{n\in\mathcal{V}}\sum_{i\in\mathcal{Q}_n}2\left(\mathbb{E}\left[f_i(\bx_i^{(0)},\bar{\bz}^{(0)})\right]-\mathbb{E}\left[f_i(\bx_i^{(T)},\bar{\bz}^{(T)})\right]\right)$.

Given the definition of $e_{x_i}^{(t)}$, suppose that $t_i$ is the element in the subsequence $\{T_i\}\subseteq\{0,1,\ldots,T-1\}$ such that $i\in\mathcal{Q}_n^{(t_i)}$ for all $t_i$ [the existence of such subsequence is guaranteed by Assumption~\ref{Asm:subset}(i)], we also notice that
\begin{align}
    \sum_{t=0}^{T-1}e_{x_i}^{(t)}
    =& \sum_{t\in\{T_i\}}\left(\frac{1}{\eta_x}\mathbb{E}[\|\bx_i^{(t+1)}-\bx_i^{(t)}\|^2]-\zeta^2\left(\frac{1}{\eta_x}+\rho_i\right)\mathbb{E}[\|\bx_i^{(t)}-\bx_i^{(t^-)}\|^2]\right)\nonumber\\
    \geq&\sum_{t\in\{T_i\}}\left(\frac{1}{\eta_x}\mathbb{E}[\|\bx_i^{(t+1)}-\bx_i^{(t)}\|^2]-\frac{\omega^2}{\eta_x}\mathbb{E}[\|\bx_i^{(t)}-\bx_i^{(t^-)}\|^2]\right) \nonumber\\
    \geq&\sum_{t\in\{T_i\}}\frac{(1-\omega^2)}{\eta_x}\mathbb{E}[\|\bx_i^{(t)}-\bx_i^{(t^-)}\|^2]-\frac{1}{\eta_x}\mathbb{E}[\|\bx_i^{(0)}-\bx_i^{(0^-)}\|^2]\\
\label{eq:append_c1}
    \geq&\sum_{t\in\{T_i\}}\frac{(1-\omega^2)}{\eta_x}\mathbb{E}[\|\bx_i^{(t)}-\bx_i^{(t^-)}\|^2],
\end{align}
where the first inequality uses Assumption~\ref{asm:stepsize}, and the last inequality is due to the initialization applied.
It follows that
\begin{align}
\label{eq:block_11}
    \sum_{t=0}^{T-1}\sum_{n\in\mathcal{V}}\sum_{i\in\mathcal{Q}_n}(e_{x_i}^{(t)}-C_{x_i}^{(t)})
    \geq& \sum_{t=0}^{T-1}\sum_{n\in\mathcal{V}}\sum_{i\in\mathcal{Q}_n}(\eta_x\tilde{e}_{x_i}^{(t^-)}-C_{x_i}^{(t)})
    = \sum_{t=0}^{T-1}\sum_{n\in\mathcal{V}}\sum_{i\in\mathcal{Q}_n}\eta_x(\tilde{e}_x^{(t^-)}-\tilde{C}_{x_i}^{(t^-)}),
\end{align}
where
\[
    \tilde{e}_{x_i}^{(t^-)} = \begin{cases}\frac{1-\omega^2}{\eta_x^2}\mathbb{E}[\|\bx_i^{(t)}-\bx_i^{(t^-)}\|^2] & i\in\mathcal{Q}_n^{(t)}\\0 & {\rm otherwise}\end{cases},~~\tilde{C}_{x_i}^{(t^-)} = \begin{cases}\frac{C_{x_i}^{(t)}}{\eta_x}& i\in\mathcal{Q}_n^{(t)}\\0, & {\rm otherwise}\end{cases}.
\]
Combining~\eqref{eq:main_ttt} and~\eqref{eq:block_11}, we obtain
\begin{align*}
    &\sum_{t=0}^{T-1}\left(\sum_{n\in\mathcal{V}}\sum_{i\in\mathcal{Q}_n}\eta_x(\tilde{e}_{x_i}^{(t^-)}-\tilde{C}_{x_i}^{(t^-)})+\eta_z\left(e_z^{(t)}-C_z^{(t)}\right)\right)
    \leq
    \Delta_T\leq \Delta^*.
\end{align*}
where $\Delta^*=\sum_{n\in\mathcal{V}}\sum_{i\in\mathcal{Q}_n}2\left(\mathbb{E}\left[f_i(\bx_i^{(0)},\bar{\bz}^{(0)})\right]-f_i^*\right)$ is a constant and $f_i^*$ denotes the minimum value for $f_i$.
The above inequality can be realigned as follows
\begin{equation}
    \frac{\eta_{\sf min}}{T}\sum_{t=0}^{T-1}\left(\sum_{n\in\mathcal{V}}\sum_{i\in\mathcal{Q}_n}\tilde{e}_{x_i}^{(t^-)}+e_z^{(t)}\right)
    \leq
    \frac{\Delta^*}{T} + \frac{\eta_{\sf max}}{T}\sum_{t=0}^{T-1}\left(\sum_{n\in\mathcal{V}}\sum_{i\in\mathcal{Q}_n}\tilde{C}_{x_i}^{(t^-)}+C_z^{(t)}\right),
\end{equation}
where $\eta_{\sf min} = \min\{\eta_z,\eta_x\}$ and $\eta_{\sf max} = \max\{\eta_z,\eta_x\}$.

Our next goal is to show that $\frac{\eta_{\sf max}}{T}\sum_{t=0}^{T-1}\left(\sum_{n\in\mathcal{V}}\sum_{i\in\mathcal{Q}_n}\tilde{C}_{x_i}^{(t^-)}+C_z^{(t)}\right)$ is bounded.
It can be verified that
\begin{align*}
    & \sum_{t=0}^{T-1}\left(\sum_{n\in\mathcal{V}}\sum_{i\in\mathcal{Q}_n}\tilde{C}_{x_i}^{(t^-)}+C_z^{(t)}\right)\\
    =& \sum_{t=0}^{T-1}\sum_{n\in\mathcal{V}}\sum_{i\in\mathcal{Q}_n}2\mathbb{E}[(\zeta D_{\mathcal{X}}+\eta_x(D_{g_i}+\gamma_i D_{\mathcal{X}}+\|\bm\delta_i^{(t)}\|))(\gamma_i\|\bar{\bz}^{(t)}-\bz_n^{(t)}\|+\|\bm\delta_i^{(t)}\|)]/\eta_x+\gamma^2|\mathcal{Q}|^2\sum_{t=0}^{T-1}\sum_{n\in\mathcal{V}}\mathbb{E}[\|\bar{\bz}^{(t)}-\bz_n^{(t)}\|^2]\\
    \leq& \sum_{t=0}^{T-1}\sum_{n\in\mathcal{V}}\sum_{i\in\mathcal{Q}_n}2\gamma(\zeta D_{\mathcal{X}}/\eta_x+D_{g_i}+\gamma D_{\mathcal{X}}+\mathbb{E}[\|\bm\delta_i^{(t)}\|])\mathbb{E}[\|\bar{\bz}^{(t)}-\bz_n^{(t)}\|]+
    \sum_{t=0}^{T-1}\sum_{n\in\mathcal{V}}\sum_{i\in\mathcal{Q}_n}2\mathbb{E}[\|\bm\delta_i^{(t)}\|^2]+\\
    &\sum_{t=0}^{T-1}\sum_{n\in\mathcal{V}}\sum_{i\in\mathcal{Q}_n}2(\zeta D_{\mathcal{X}}/\eta_x+D_{g_i}+\gamma D_{\mathcal{X}})\mathbb{E}[\|\bm\delta_i^{(t)}\|]+
    \gamma^2|\mathcal{Q}|^2\sum_{t=0}^{T-1}\sum_{n\in\mathcal{V}}\mathbb{E}[\|\bar{\bz}^{(t)}-\bz_n^{(t)}\|^2]\\
    \leq&2\gamma|\mathcal{V}||\mathcal{Q}|(\zeta D_{\mathcal{X}}/\eta_x+D_{g_i}+\gamma D_{\mathcal{X}}+\sigma/\sqrt{R})(C_1+\eta_zTC_2)+2\sigma^2T/R+2T|\mathcal{V}||\mathcal{Q}|(\zeta D_{\mathcal{X}}/\eta_x+D_{g_i}+\gamma D_{\mathcal{X}})\sigma/\sqrt{R}\\
    &+\gamma^2|\mathcal{V}||\mathcal{Q}|^2(C_3+\eta_zTC_4+\eta_zC_5+\eta_z^2TC_6)\\
    =& C_7 + T(\eta_zC_8 + \eta_z^2C_{9} + 2\sigma^2 /R + \sigma C_{10}/\sqrt{R}),
\end{align*}
where the second inequality uses Fact~\ref{fact:stochastic error}, Lemma~\ref{lemma:z_consensus} and the fact that $\mathbb{E}\left[\|\bm v\|\right]\leq \sqrt{\mathbb{E}\left[\left\|\bv\right\|^2\right]}$ for any random variable $\bv$, and $C_1$-$C_{11}$ are introduced in Lemma~\ref{lemma:accumulation}.
Recalling that we apply $\eta_z=d/\sqrt{T}$, it follows that
\begin{align*}
    \frac{\eta_{\sf min}}{T}\sum_{t=0}^{T-1}\left(\sum_{n\in\mathcal{V}}\sum_{i\in\mathcal{Q}_n}\tilde{e}_{x_i}^{(t^-)}+e_z^{(t)}\right)
    \leq&
    \frac{\Delta^*+\eta_{\sf max}(C_7+d^2C_9)}{T}+ \frac{\eta_{\sf max}dC_8}{\sqrt{T}} + \frac{\eta_{\sf max}( 2\sigma^2 /\sqrt{R} + \sigma C_{10})}{\sqrt{R}}.
\end{align*}
Substituting the definitions of $\tilde{e}_{x_i}^{(t^-)}$ and $e_z^{(t)}$ and since $\sqrt{T} \geq \frac{\Delta^*+\eta_{\sf max}(C_7+d^2C_9)}{\eta_{\sf max}dC_8}$,
it can then be easily checked that
\begin{align*}
    &\frac{1}{T}\sum_{t=0}^{T-1}\mathbb{E}\left[\left\|\sum_{n\in\mathcal{V}}\sum_{i\in\mathcal{Q}_n}\nabla_{\bz_n}f_i(\bx_i^{(t+1)},\bar{\bz}^{(t)})\right\|^2\right] \leq \frac{B_1}{\sqrt{T}}+\frac{B_2}{\sqrt{R}},\\
    &\frac{1}{T}\sum_{t=0}^{T-1}\mathbb{E}\left[\frac{1}{\eta_x^2}\left\|\bx_i^{(t)}-\bx_i^{(t-1)}\right\|^2\right] = \frac{1}{T}\sum_{t_i\in\{T_i\}}\mathbb{E}\left[\frac{1}{\eta_x^2}\left\|\bx_i^{(t_i)}-\bx_i^{(t_i^-)}\right\|^2\right] \leq \frac{B_1}{\sqrt{T}}+\frac{B_2}{\sqrt{R}}.
\end{align*}
We complete the proof.

\end{appendix}

\bibliographystyle{IEEEtran}
\bibliography{refs}

%


\end{document}